\documentclass[12pt, a4paper]{article}
\usepackage{amssymb,amsmath,mathtools,xcolor,graphicx,xspace,colortbl,ragged2e,rotating}
\usepackage{amsfonts}
\usepackage{amsmath,amssymb,setspace}
\usepackage{amssymb,amsmath,mathtools,xcolor,graphicx,xspace,colortbl,ragged2e,rotating}
\usepackage{graphics,color}
\usepackage{makeidx}
\usepackage[english]{babel}
\usepackage{float}
\usepackage{caption}
\usepackage{subcaption}
\usepackage[lofdepth,lotdepth]{subfig}
\usepackage{geometry}

\graphicspath{{JW250322_graphics/}{JW250322_tcache/}{JW250322_gcache/}}
\DeclareGraphicsExtensions{.pdf,.eps,.ps,.png,.jpg,.jpeg}
\graphicspath{{JW280122_graphics/}{JW280122_tcache/}{JW280122_gcache/}}
\DeclareGraphicsExtensions{.pdf,.eps,.ps,.png,.jpg,.jpeg}
\graphicspath{{JW240122_modified_graphics/}{JW240122_modified_tcache/}{JW240122_modified_gcache/}}
\DeclareGraphicsExtensions{.pdf,.eps,.ps,.png,.jpg,.jpeg}
\graphicspath{{JW240122_modified_graphics/}{JW240122_modified_tcache/}{JW240122_modified_gcache/}}
\DeclareGraphicsExtensions{.pdf,.eps,.ps,.png,.jpg,.jpeg}
\graphicspath{{JW240122_modified_graphics/}{JW240122_modified_tcache/}{JW240122_modified_gcache/}}
\DeclareGraphicsExtensions{.pdf,.eps,.ps,.png,.jpg,.jpeg}
\graphicspath{{JW190122_modified_graphics/}{JW190122_modified_tcache/}{JW190122_modified_gcache/}}
\DeclareGraphicsExtensions{.pdf,.eps,.ps,.png,.jpg,.jpeg}
\newtheorem{theorem}{Theorem}
\newtheorem{proposition}{Proposition}
\newtheorem{definition}{Definition}

\newtheorem{lemma}{Lemma}

\usepackage{changepage}

\begin{document}

\title{{\large {Endogenous Clustering and Analogy-Based Expectation
Equilibrium}\thanks{%
Previous versions have circulated with the title \textquotedblleft
Calibrated Clustering and Analogy-Based Expectation
Equilibrium\textquotedblright . We thank three reviewers, the editor (Andrea
Galeotti) as well as Evan Friedman, Fr\'{e}d\'{e}ric Koessler, Giacomo
Lanzani, David Levine, Gilat Levy, Olivier Tercieux and Bob Gibbons. We also
thank audiences at seminars at PSE, Paris 1, Berlin, Vienna, Royal Holloway,
Warwick university and at the Parisian Seminar of Game Theory, Bounded
Rationality: Theory and Experiments workshop in Tel Aviv, ESSET 2023, AMES
2023 in Singapore, Summer School of Econometric Society in Economic Theory
2023, LSE-UCL Economic Theory Workshop 2023, BSE Summer Forum 2024, and EEA
2024 for comments. We thank the support of the EUR grant ANR-17-EURE-0001
and we thank the European Research Council for funding (grant n$%
{{}^\circ}%
$ 742816). Giacomo Weber thanks the Italian Ministry of University and Research for funding under the PRIN 2022 program (Prot. 20228XTY79).}}}
\author{{\large {Philippe Jehiel}\thanks{%
PSE, 48 boulevard Jourdan, 75014 Paris, France and University College
London; jehiel@enpc.fr} {and Giacomo Weber}\thanks{%
PSE (EHESS), 48 boulevard Jourdan, 75014 Paris, France and University of
Bologna, Italy; giacomo.weber@psemail.eu }}}
\maketitle

\begin{abstract}
Normal-form two-player games are categorized by players into K analogy
classes so as to minimize the prediction error about the behavior of the
opponent. This results in Clustered Analogy-Based Expectation Equilibria in
which strategies are analogy-based expectation equilibria given the analogy
partitions and analogy partitions minimize the prediction errors given the
strategies. We distinguish between environments with self-repelling analogy
partitions in which some mixing over partitions is required and environments
with self-attractive partitions in which several analogy partitions can
arise, thereby suggesting new channels of belief heterogeneity and
equilibrium multiplicity. Various economic applications are discussed.

\noindent \textbf{Keywords}: Analogy-based Expectation Equilibrium,
Prototype theory, K-means clustering. \newline
\textbf{JEL Classification Numbers:} D44, D82, D90
\end{abstract}

\newpage
\section{Introduction}

Many economists recognize that the rational expectation hypothesis that is
central in solution concepts such as the Nash equilibrium is very demanding,
especially in complex multi-agent environments involving lots of different
situations (games, states or nodes, depending on the application). Several
approaches have been proposed to relax it. When the concern with the
hypothesis is that there are too many situations for players to fine tune a
specific expectation for each such situation, a natural approach consists in
allowing players to lump together situations into just a few categories, and
only require that players form expectations about the aggregate play in each
category.

The analogy-based expectation equilibrium (Jehiel, 2005) is a solution
concept that has been proposed to deal with this. In addition to the usual
primitives describing a game form, players are also endowed with analogy
partitions, which are player-specific ways of partitioning situations or
contingencies in the grand game. In equilibrium, the expectations in each
analogy class correctly represent the aggregate behavior in the class, and
players best-respond as if the behavior in every element of an analogy class
matched the expectation about the aggregate play in the corresponding
analogy class. This approach has been developed and applied to a variety of
settings, but in almost all these developments, the analogy partitions are
taken as exogenous (see Jehiel (2022) for a recent exposition of this strand
of literature).

From a different perspective, psychologists have long recognized the use of
categories to facilitate decision making and predictions (see, in
particular, Anderson (1991) on predictions). For leading approaches in
psychology (see in particular Rosch (1978)), categories bundle distinct
objects that are viewed as sufficiently similar to warrant a similar
treatment. Each category is associated with a prototype that can be viewed
as a representative object,\footnote{%
The representative object is possibly fictitious (e.g. mean or mode of
objects in the category). This corresponds to the view developed in the
literature on prototype theory (see Rosch (1978) for an early account). An
alternative approach is that of exemplar theory (Medin and Schaffer (1978),
Nosofsky (1986)) in which only real existing objects are considered to
describe the category. Such an alternative approach would introduce an
element of stochasticity in the choice of representative exemplar that is
somehow orthogonal to the focus of the present study, hence our primary
reference to the prototype approach.} and categories are defined so that
objects are assigned to the category with nearest prototype (see Posner and
Keele (1968) or Reed (1972)).

From yet another perspective, the K-means clustering technique considered in
Machine Learning has been widely used as a method to categorize large sets
of exogenously given datapoints into a pre-specified number K of clusters
(see Steinhaus (1957), Lloyd (1957) and MacQueen (1967)). From that
perspective, datapoints are the primitives, and the clustering problem
consists in partitioning the datapoints into K clusters with representative
points for each cluster defined so that the original datapoints are best
approximated by the representative points in their cluster. Solving the
clustering problem (defined as deriving the variance-minimizing
categorization, say) is hard (NP-complete), and practitioners most of the
time rely on a simple algorithm to approximate its solution (see the end of
Subsection 2.3 for a succinct exposition of the algorithm).

In this paper, we propose endogenizing the choice of partitions in the
analogy-based expectation setting on the basis of the above basic
principles. Specifically, we consider a strategic environment consisting of
different normal form two-player games drawn by nature according to some
prior distribution where we have in mind that the various games are played
at many different times by many different subjects. In each of the normal
form games $\omega \in \Omega $, player $j=1,2$ has the same action set $%
A_{j}$. An analogy partition for player $i$ takes the form of a partition of
the set of games $\Omega $, which is used by player $i$ to assess the
behavior of player $j$ in the various games. The data points accessible by
players consist of the empirical frequencies of past play of the subjects
assigned to the role of the opponent in the various games. To make sense of
these data points, a subject assigned to the role of player $i$ is viewed as
clustering these data points into an exogenously given (typically small)
number $K$ of categories where $K$ can be related to the number of items
human beings can remember in short-term memory (see Miller (1956) for
pioneering research on this).\footnote{%
Even if the exact formulation in terms of number of categories is specific
to our setting, at some broad level, one can relate the bound $K$ in our
approach to the maximum number of states considered in automaton theory (see
in particular Neyman (1985) or the discussion in Rubinstein (1986)) or to
the maximum number of clauses that can be considered by an agent in a
contract setting (Jakobsen (2020)).} At the end of the clustering stage,
each game is assigned to the cluster with nearest opponent's representative
behavior and the representative behavior in a cluster is identified with the
mean behavior of the opponent across the games assigned to the cluster.%
\footnote{%
More precisely, in the learning section, we consider either that the subject
does the clustering himself as in the ML literature or that he inherits
representative points from some subject in an earlier generation (assigns
the datapoints to the nearest representative point and use as final
representative points the means in each cluster).} In a later stage, the
relevant game $\omega $ becomes known to the subject, the player then
identifies the behavior of his opponent in this game $\omega $ with the
representative expectation in the corresponding cluster, and best-responds
to it. This in turn generates new data points, and we are interested in the
steady states - referred to as clustered analogy-based expectation
equilibria - generated by such dynamic processes.

Roughly, the clustered analogy-based expectation equilibria (CABEE) can be
described as profiles of analogy partitions and strategies such that i)
given the analogy partitions, players' strategies form an analogy-based
expectation equilibrium and ii) given the strategies, clustering leads
players to adopt the analogy partitions considered in steady state.

A key observation is that, in some cases, a steady state with a single
analogy partition for each player does not exist. This follows because
unlike in the usual clustering problem, there is here an extra endogeneity
of the dataset. A change in analogy partitions may affect the adopted
strategies through the working of the analogy-based expectation equilibrium,
which in turn may affect how clustering is done. This will be first
illustrated with a simple example involving three matching pennies games.

This observation leads us to extend the basic definition of CABEE to allow
for distributions over analogy partitions defined so that each analogy
partition in the support is required to solve (either locally or globally)
the clustering problem for that player and the strategies, now parametrized
by the chosen analogy partition, satisfy the requirements of the
analogy-based expectation equilibrium appropriately extended to cope with
distributions over analogy partitions. We refer to such an extension as a
clustered distributional analogy-based expectation equilibrium (CD-ABEE).

We show that in finite environments (i.e. environments such that there are
finitely many normal form games and finitely many actions for each player),
there always exists at least one CD-ABEE. We also provide two learning
dynamic models that give rise to CD-ABEE as steady states. In one of them,
as informally described above, the learning subjects first cluster the raw
data in each game before they are told (typically with a lag) which game
applies to them. We note that when several analogy partitions are required
in steady state, it implies that different subjects assigned to the role of
the same player would end up categorizing the various games differently,
despite being exposed to the same objective datasets. In the second learning
model, subjects are told categories and representative points by their
parents, then they play accordingly in the various games. In a later stage,
subjects arrange newly observed behaviors in these games in new categories
and pass these as well as the corresponding representative points to their
off-springs.

We next consider various applications. A common theme that we consider
throughout is whether the environment is such that starting from any
candidate analogy partition profile, we obtain behaviors (through the
machinery of ABEE) that would lead to the re-categorization of at least one
game (through the machinery of clustering) or else whether the environment
is such that starting from several (sometimes many) analogy partition
profiles, we obtain (ABEE) behaviors that in turn lead to the same partition
profile from the clustering perspective. We say that the environment has
self-attractive analogy partitions in the latter case and self-repelling
ones in the former. Environments with self-attractive analogy partitions
suggest for such environments a novel channel through which different
societies may settle down on different behaviors and different models of
expectation formation. By contrast, environments with self-repelling analogy
partitions lead to CD-ABEE (with mixed distributions of analogy partitions)
and can be viewed as suggesting in such settings a new channel of
heterogeneity in behaviors and expectation formation within societies.

We first provide an illustration of an environment with self-attractive
analogy partitions in the context of beauty contest games in which players
care both about being close to a fundamental value (which parametrizes the
game) and being close to what the opponent is expected to be choosing. We
show that the set of CABEE expands as the concern for coordination with the
opponent increases. We also observe that when the coordination motive is
sufficiently strong, virtually all analogy partitions of the fundamental
value can be used to construct a CABEE, and as a result many different
strategies can arise in a CABEE. The latter result illustrates in a very
extreme way a case of self-attractive analogy partitions.

We next illustrate the possibility of self-repelling analogy partitions in
the context of a monitoring game involving an employer who has to choose
whether or not to exert some control and a worker who can be one of three
possible types (which defines the game) and has to decide his effort
attitude. When the employer can use two categories and only one minority
type is responsive to the monitoring technology, we observe that
heterogeneous categories are needed, thereby shedding new light on the study
of discrimination in such applications.

Finally, we consider families of games with linear best-responses
parametrized by the magnitude of the impact of opponent's action on the
best-response. We analyze separately the case of strategic complements and
the case of strategic substitutes allowing us to cover applications such as
Bertrand or Cournot duopoly with product differentiation, linear demand and
constant marginal costs. We show that the strategic complements case
corresponds to an environment with self-attractive analogy partitions and
the strategic substitute case corresponds to an environment with
self-repelling analogy partitions.

In the rest of the paper, we develop the framework (solution concepts,
existence results, learning foundation) in Section 2. We develop
applications in Section 3. We conclude in Section 4.

\subsection{Related Literature}

This paper belongs to a growing literature in behavioral game theory,
proposing new forms of equilibrium designed to capture various aspects of
misperceptions or cognitive limitations. While some papers in this strand
posit some misperceptions of the players and propose a corresponding notion
of equilibrium (see Eyster-Rabin (2005) on misperceptions about how private
information affects behavior, Spiegler (2016) on misperceptions on the
causality links between variables of interest or Esponda-Pouzo (2016) for a
more abstract and general formulation of misspecifications), other papers
motivate their equilibrium approach by the difficulty players may face when
trying to understand or learn how their environment behaves (see Jehiel
(1995) on limited horizon forecasts, Osborne-Rubinstein (1998) on sampling
equilibrium, Jehiel (2005) on analogical reasoning or Jehiel-Samet (2007) on
coarse reinforcement learning). Our paper has a motivation more in line with
the latter, but it adds structure on the coarsening of the learning based on
insights or techniques borrowed from psychology and/or machine learning
(which the previous literature just mentioned did not consider).\footnote{%
Our approach can be related to the self-confirming equilibrium (Battigalli,
1987) to the extent that our players form their expectations based on coarse
statistics that describe only partially others' behaviors. A key difference
is that the coarse statistics used by our players are not exogenously given
as in the self-confirming equilibrium, but they are endogenously determined
from the actual behaviors through the clustering machinery (see Jehiel
(2022) for complementary discussion of the link of the analogy-based
expectation equilibrium to the self-confirming equilibrium).}

This paper also relates to papers dealing with coarse or categorical
thinking in decision-making settings (see, in particular, Fryer and Jackson
(2008) for such a model used to analyze stereotypes or discrimination, Peski
(2011) for establishing the optimality of categorical reasoning in symmetric
settings or Al-Najjar and Pai (2014) and Mohlin (2014) for models
establishing the superiority of using not too fine categories in an attempt
to mitigate overfitting or balance the bias-variance trade-off). Our paper
considers a clustering technique not discussed in those papers, and due to
the strategic character of our environment the data-generating process in
our setting is itself affected by the categorization, which these papers did
not consider.

Finally, a contemporaneous and alternative approach to categorization in the
context of the analogy-based expectation equilibrium is that introduced in
Jehiel and Mohlin (2023) who propose putting structure on the analogy
partitions based on the bias-variance trade-off where an exogenously given
notion of distance between the various nodes or information sets is
considered by the players. This is a different approach to categorization
than the one considered in this paper, and it is more impactful in contexts
such as extensive form games in which the behaviors in some nodes cannot be
observed without trembles. In the context considered here with a set of
normal form games (all arising with positive probability) it would lead to
consider the finest analogy partition and accordingly the Nash Equilibrium
would arise, in sharp contrast with the findings developed below.\footnote{%
Some other papers consider categorization in games (see in particular
Samuelson (2001), Mengel (2012) or Gibbons et al. (2021)) with the view that
the strategy should be measurable with respect to the categorization. This
is somewhat different from the expectation perspective adopted here. It may
be mentioned that Gibbons et al. (2021) discuss the possibility that some
third party could influence how categorizations are chosen, which differs
from our perspective that views the categories as being chosen by the
players themselves.}

\section{Theoretical setup}

\subsection{Strategic environment}

We consider a finite number of normal form games indexed by $\omega \in
\Omega $ where game $\omega $ is chosen (by Nature) with probability $%
p(\omega )$. To simplify the exposition, we restrict attention to games with
two players $i=1,2$, and we refer to player $j$ as the player other than $i$.%
\footnote{%
The framework, solution concept and analysis extend in a straightforward way
to the case of more than two players.} In every game $\omega $, the action
space of player $i$ is the same and denoted by $A_{i}$. It is assumed in
this part to be finite. The payoff of player $i$ in game $\omega $ is
described by a von Neumann-Morgenstern utility where $u_{i}(a_{i},a_{j},%
\omega )$ denotes the payoff obtained by player $i$ in game $\omega $ if
player $i$ chooses $a_{i}\in A_{i}$ and player $j$ chooses $a_{j}\in A_{j}$.
Let $p_{i}\in \Delta A_{i}$ denote a probability distribution over $A_{i}$
for $i=1,2$. With some abuse of notation, we let: 
\begin{equation*}
u_{i}(p_{i},p_{j},\omega )={\displaystyle\sum_{a_{i},a_{j}}}%
p_{i}(a_{i})p_{j}(a_{j})u_{i}(a_{i},a_{j},\omega )
\end{equation*}%
denote the expected utility obtained by player $i$ in game $\omega $ when
players $i$ and $j$ play according to $p_{i}$ and $p_{j}$, respectively.

We assume that players observe the game $\omega $ they are in when choosing
their actions. A strategy for player $i$ is denoted $\sigma _{i}=\left(
\sigma _{i}(\omega )\right) _{\omega \in \Omega }$ where $\sigma _{i}(\omega
)\in \Delta A_{i}$ denotes the (possibly mixed) strategy employed by player $%
i$ in game $\omega $. The set of player $i$'s strategies is denoted $\Sigma
_{i}$, and we let $\Sigma =\Sigma _{i}\times \Sigma _{j}$.

A Nash equilibrium is a strategy profile $\sigma =(\sigma _{i} ,\sigma _{j})
\in \Sigma $ such that for every player $i$, $\omega \in \Omega $, and $%
p_{i} \in \Delta A_{i}$, 
\begin{equation*}
u_{i} (\sigma _{i} (\omega ) ,\sigma _{j} (\omega ) ,\omega ) \geq u_{i}
(p_{i} ,\sigma _{j} (\omega ) ,\omega )\text{.}
\end{equation*}

\subsection{Analogy-based expectation equilibrium}

Players are not viewed as being able to know or learn the strategy of their
opponent for each game $\omega $ separately as implicitly required in Nash
equilibrium. Maybe because there are too many games $\omega $, they are
assumed to learn the strategy of their opponent only in aggregate over
collections of games, referred to as analogy classes. Throughout the paper,
we impose that player $i$ considers (at most) $K_{i}$ different analogy
classes, where $K_{i}$ is kept fixed. Such a bound can be the result of
constraints on memory (see the introduction), and we will have in mind that $%
K_{i}$ is no greater and typically (possibly much) smaller than $\left\vert
\Omega \right\vert $, the number of possible normal form games. We refer to $%
\mathcal{K}_{i}$ as the set of partitions of $\Omega $ with $K_{i}$
elements. Formally, considering for now the case of a single analogy
partition for each player $i$, we let $An_{i}=\left\{ \alpha
_{i}^{1},\ldots, \alpha _{i}^{K_{i}}\right\} $ denote the analogy partition
of player $i$. It is a partition of the set $\Omega $ of games with $K_{i}$
classes, hence an element of $\mathcal{K}_{i}$. For each $\omega \in \Omega $%
, we let $\alpha _{i}(\omega )$ denote the (unique) analogy class to which $%
\omega $ belongs.

$\beta _{i}(\alpha _{i})\in \Delta A_{j}$ will refer to the (analogy-based)
expectation of player $i$ in the analogy class $\alpha _{i}$. It represents
the aggregate behavior of player $j$ across the various games $\omega \;$in $%
\alpha _{i}$. We say that $\beta _{i}$ is \textit{consistent} with $\sigma
_{j}$ whenever for all $\alpha _{i}\in An_{i}$, 
\begin{equation*}
\beta _{i}(\alpha _{i})={\displaystyle\sum_{\omega \in \alpha _{i}}}p(\omega
)\sigma _{j}(\omega )/{\displaystyle\sum_{\omega \in \alpha _{i}}}p(\omega )%
\text{.}
\end{equation*}

In other words, consistency means that the analogy-based expectations
correctly represent the aggregate behaviors in each analogy class when the
play is governed\ by $\sigma $. We say that $\sigma _{i}$ is a \textit{%
best-response} to $\beta _{i}$ whenever for all $\omega \in \Omega $ and all 
$p_{i}\in \Delta A_{i}$, 
\begin{equation*}
u_{i}(\sigma _{i}(\omega ),\beta _{i}(\alpha _{i}(\omega )),\omega )\geq
u_{i}(p_{i},\beta _{i}(\alpha _{i}(\omega )),\omega )\text{.}
\end{equation*}%
In other words, player $i$ best-responds in $\omega $ as if player $j$
played according to $\beta _{i}(\alpha _{i}(\omega ))$ in this game, which
can be viewed as the simplest representation of player $j$'s strategy given
the coarse knowledge provided by $\beta _{i}$.

\begin{definition}
Given the strategic environment and the profile of analogy partitions $%
An=(An_{i},An_{j})$, $\sigma $ is an analogy-based expectation equilibrium
(ABEE) if and only if there exists a profile of analogy-based expectations $%
\beta =(\beta _{i},\beta _{j})$\ such that for each player $i$ (i) $\sigma
_{i}$ is a best-response to $\beta _{i}$ and (ii) $\beta _{i}$ is consistent
with $\sigma _{j}$.
\end{definition}

This concept has been introduced with greater generality in Jehiel (2005)
(allowing for multiple stages and more than two players) and in Jehiel and
Koessler (2008) (allowing for private information).\footnote{%
See Jehiel (2022) for a definition in a setting covering both aspects and
allowing for distributions over analogy partitions.} We have chosen a
simpler environment to focus on the choice of analogy partitions, which is
the main concern of this paper.

\subsection{Clustering}

Psychologists have long recognized the use of categories to facilitate
decision making and predictions (see, in particular, Anderson (1991) on
predictions). A categorization bundles distinct objects into groups or
categories, whose members are viewed as sufficiently similar to warrant a
similar treatment. In each category, there is a prototype that can be viewed
as a representative element in the category (say the mean or the mode, see
Rosch (1978)), and categories are defined so that objects are assigned to
the category with nearest prototype (Posner and Keele (1968) or Reed
(1972)). Another approach to categorization would require that objects are
categorized so as to minimize some measure of dispersion such as the
variance (or the relative entropy when objects can be identified with
probability distributions, as in our environment).

In our framework, the objects considered by a subject assigned to the role
of player $i$ consist of the frequencies of choices of the subjects assigned
to the role of player $j$ in the various normal form games $\omega $ (this
is made explicit when describing the learning environment, see section 2.6).
That is, for player $i$, the objects are $\left\{ (\omega ,\sigma
_{j}(\omega ))\right\} _{\omega \in \Omega }$. An obvious attribute of $%
(\omega ,\sigma _{j}(\omega ))$ is the distribution of opponent's behavior
as described by $\sigma _{j}(\omega )$, and we will assume that player $i$
focuses on that attribute when choosing his categorization. This will in
turn lead the chosen categorization to minimize (either locally or globally)
the prediction error about the play of the opponent. This feature, we
believe, would naturally be regarded as desirable by player $i$, as
predicting the opponent play is the only thing player $i$ cares about to
determine his best-response.\footnote{%
Considering extra attributes in relation to $\omega $ would not raise
conceptual difficulties, and our main insights would carry over as long as
some positive weight is given to this attribute. See further discussion in
subsection 2.6.}

We first introduce a measure of closeness in the space of distributions over
actions. Formally, for three distributions of player $j$'s play $p_{j}$, $%
p_{j}^{\prime }$ and $p_{j}^{\prime \prime }$ in $\Delta A_{j}$, we say that 
$p_{j}$ is less well approximated by $p_{j}^{\prime }$ than by $%
p_{j}^{\prime \prime }$ whenever $d(p_{j},p_{j}^{\prime
})>d(p_{j},p_{j}^{\prime \prime })$ where $d$ is a divergence function that
will either be the square of the Euclidean distance $d(p_{j},p_{j}^{\prime
}\,)=\left\Vert p_{j}-p_{j}^{\prime }\right\Vert ^{2}$ (as defined over $%
\left( \Delta A_{j}\right) ^{2}$) or the Kullback-Leibler divergence applied
to distributions $d(p_{j},p_{j}^{\prime })={\displaystyle\sum_{a_{j}}}%
p_{j}(a_{j})\ln \frac{p_{j}^{\,^{\,}}\left( a_{j}\right) }{p_{j}^{\prime
}\left( a_{j}\right) ^{\,}}$.\footnote{%
In some of the applications to be developed later, there is a natural
topology in the action space, the best-responses are linear in the mean
action of the opponent, and we identify the strategy with the mean action
that it induces. Accordingly, for $d$, we consider $d(p_{j},p_{j}^{\prime
})=\left( E(p_{j})-E(p_{j}^{\prime })\right) ^{2}$. Note that for $%
\alpha\subset\Omega\subseteq\mathbb{R}$ with density $f$ and for a function $%
g:\Omega\rightarrow\Delta\mathbb{R}$, we have that $E\left(g(\omega)\lvert%
\omega\in\alpha\right)=\arg\min_{q\in\mathbb{R}}\int_{\alpha}\left(E(g(%
\omega))-q\right)^2 f(\omega\lvert\alpha)d\omega$.}

We next define two notions of clustering relating the analogy partition of
player $i$ to the strategy of player $j$.

\begin{definition}
A partition $An_{i}$ of $\Omega $ is locally clustered with respect to $%
\sigma _{j}$ iff for all classes $\alpha _{i}$, $\alpha _{i}^{\prime }$ of $%
An_{i}$ and every $\omega \in \alpha _{i}$, 
\begin{equation*}
d(\sigma _{j}(\omega ),\beta _{i}(\alpha _{i}))\leq d(\sigma _{j}(\omega
),\beta _{i}(\alpha _{i}^{\prime }))\text{.}
\end{equation*}%
It is globally clustered with respect to $\sigma _{j}$ iff $An_{i}\ $belongs
to 
\begin{equation}
\arg \min_{P_{i}\in \mathcal{K}_{i}}{\displaystyle\sum_{c_{i}\in P_{i}}}%
p(c_{i}){\displaystyle\sum_{\omega \in c_{i}}}p(\omega \mid c_{i})d(\sigma
_{j}(\omega ),\beta _{i}(c_{i}))  \label{OBJ}
\end{equation}%
where, for all $c\subseteq \Omega $, $\beta _{i}(c)=$ ${\displaystyle%
\sum_{\omega \in c}}p(\omega \mid c)\sigma _{j}(\omega )$.
\end{definition}

In the above definition, $\beta _{i}(c)$ is viewed as the prototype in
category $c$ and it is defined as the mean of the elements assigned to $c$.
Local clustering retains the idea that objects should be assigned to the
category with nearest prototype. Global clustering on the other hand
requires that categorizations are chosen to minimize dispersion as measured
by (\ref{OBJ}). When $d$ is the square of the Euclidean distance, the
measure of dispersion corresponds to variance. When $d$ is the the
Kullback-Leibler divergence, the measure of dispersion corresponds to
relative entropy. We note that in both cases, choosing the prototype to be
the mean of the elements assigned to the category is optimal (in the sense
of minimizing the dispersion measure), thereby giving a further argument for
using the mean as the prototype in our context.\footnote{%
Formally, for $i=1,2$ and any subset $\alpha _{i}$ of $\Omega $, let $d$ be
either the square of the Euclidean distance or the Kullback-Leibler
divergence, we have that 
\begin{equation*}
{\displaystyle\sum_{\omega \in \alpha _{i}}}p(\omega \mid \alpha _{i})\sigma
_{j}(\omega )=\arg \min_{q\in \Delta A_{j}}{\displaystyle\sum_{\omega \in
\alpha _{i}}}p(\omega \mid \alpha _{i})d(\sigma _{j}(\omega ),q).
\end{equation*}%
This property is central to establish the convergence of the K-means
algorithm and it holds for a divergence if and only if this is a Bregman
divergence, as shown by Banerjee et al. (2005). We regard the relative
entropy and the variance as the most commonly used measures of dispersion in
the class of Bregman divergences.}

It should be stressed (and as formally established in Lemma \ref%
{Loc_implies_glob} in the Online Appendix B) that global clustering implies
local clustering for the $d$ as specified above. As it turns out, the local
clustering conditions can be viewed as the first order conditions for the
minimization problem (\ref{OBJ}).\footnote{%
This conclusion holds as long as $d$ is a Bregman divergence.}

\bigskip \textbf{Link to K-means clustering.}

In Machine Learning, a very popular way to categorize objects is based on
the K-means clustering algorithm (Steinhaus (1957), Lloyd (1957) and
MacQueen (1967)). The objective of clustering is also to minimize variance
(or relative entropy) as considered in our global clustering approach, but
this is known to be NP hard in computer science. Practitioners instead use
the following algorithm. Representative points are initially drawn, then at
each subsequent iteration of the algorithm, first points are allocated to
the cluster with closest representative, then, a new representative point,
identified with the mean of the points allocated to the cluster, is
determined in each cluster. Such an algorithm can be shown to converge, and
it turns out that it necessarily converges to a local minimizer of the
dispersion measure. Moreover, all local minimizers satisfy the local
clustering condition requiring that points are assigned to the cluster with
closest representative point.

\subsection{Clustered analogy-based expectation equilibrium}

Combining Definitions 1 and 2 yield:

\begin{definition}
A pair $(\sigma ,\,An)$ of strategy profile $\sigma =(\sigma _{i},\sigma
_{j})$ and analogy partition profile $An=(An_{i},An_{j})\in \mathcal{K}%
_{i}\times \mathcal{K}_{j}$ is a locally (resp. globally) clustered
analogy-based expectation equilibrium iff (i) $\sigma $ is an analogy-based
expectation equilibrium given $An$ and (ii) for each player $i$, $An_{i}$ is
locally (resp. globally) clustered with respect to $\sigma _{j}$.
\end{definition}

The more interesting and novel aspect in this definition is the fixed point
element linking analogy partitions to strategies and vice versa. With
respect to the previous papers (using the ABEE framework), it suggests a way
to endogenize the analogy partitions (given the numbers $K_{i}$ and $K_{j}$
of allowed analogy classes). With respect to the clustering literature, the
novel aspect is that the set of points to be clustered $\{\left(
\omega,\sigma _{j}(\omega )\right) \}_{\omega \in \Omega }$ for player $i$
is itself possibly influenced by the shape of the clustering, as captured by
the analogy-based expectation equilibrium.

When some player $i$ has a dominant strategy in all games $\omega \in \Omega 
$, there always exists a (locally or globally) clustered ABEE because player 
$j$'s analogy partition can simply be obtained by clustering on the
exogenous dataset given by $i$'s dominant strategies. Also, if the number of
games $\omega $ with different Nash equilibrium strategies is no larger than 
$K_{i}$ for each player $i$, a (globally) clustered ABEE is readily obtained
by requiring that $i$ plays the corresponding Nash equilibrium strategy and
games are bundled in the same analogy class only when the opponent's
strategies coincide in such games.\footnote{%
Observe that whenever all the objects in a category coincide exactly with
the prototype, then the correct expectations would lead to play optimally.
This would not be so in general if the clustering were based on alternative
attributes (say, player $i$'s own payoff structure), thereby giving a
normative appeal to the use of opponent's behavior as the main attribute for
clustering purpose.}

But, in general, a clustered analogy-based expectation equilibrium may fail
to exist. We illustrate this in a context with three matching penny games
with different parameter values assuming that one of the players can use
only two categories.

\textbf{Example 1.} Let $x=a,b,c$, with $0<a<b<c<2$. The following three
games are played, each with probability $\frac{1}{3}$. The corresponding
payoff matrices are given by: 
\begin{equation*}
\begin{array}{ccc}
G _{x} & L & R \\ 
U & (1+x,0) & (0,1) \\ 
D & (0,1) & (1,0)%
\end{array}%
\end{equation*}

\begin{proposition}
Assume that $K_{1}=2$ and $K_{2}=3$, and $d$ is the square of the Euclidean
distance in the space of probability distributions on $\left\{ L,R\right\} $%
. There is no Clustered ABEE.
\end{proposition}

Roughly, Proposition 1 can be understood as follows. Matching pennies games
are such that the Nash equilibrium involves some mixing. When the Row player
puts two games $x$ and $x^{\prime }$ in the same analogy class, she can be
mixing in at most one of these games (this is so because the incentives of
Row are different in the two games and Row makes the same expectation about
Column in both $x$ and $x^{\prime }$ when these belong to the same analogy
class). This in turn induces some polarization of the behaviors of both
players in the two games $x$ and $x^{\prime }$, which when $a,b$ and $c$ are
not too far apart leads the Row player to re-categorize one of the two
bundled games $x$ or $x^{\prime }$ with the left alone game $x^{\prime
\prime }\neq x,x^{\prime }$ at the clustering stage. Details appear in
Appendix.

\subsection{Clustered distributional analogy-based expectation equilibrium}

We address the existence issue by allowing players to use heterogeneous
analogy partitions. Formally, the analogy partition $An_{i}$ of player $i$
may now take different realizations in $\mathcal{K}_{i}$, and we refer to $%
\lambda _{i}$ as the distribution of $An_{i}$ over $\mathcal{K}_{i}$. The
distributions of analogy partitions of the two players are viewed as
independent of one another (formalizing a random assignment assumption, see
below the section on learning). We refer to $\lambda =(\lambda _{i},\lambda
_{j})$ as the profile of these distributions, and we let $\Lambda =\Delta 
\mathcal{K}_{i}\times \Delta \mathcal{K}_{j}$ be the set of $(\lambda
_{i},\lambda _{j})$. For each analogy partition $An_{i}$ of player $i$ in
the support of $\lambda _{i}$ referred to as $Supp\lambda _{i}$, we let $%
\sigma _{i}(\cdot \mid An_{i}):\Omega \rightarrow \Delta A_{i}$ refer to the
mapping describing how player $i$ with analogy partition $An_{i}$ behaves in
the various games $\omega \in \Omega $. We refer to $\sigma _{i}=(\sigma
_{i}(\cdot \mid An_{i}))_{An_{i}\in\mathcal{K}_i}$ as player $i$'s strategy,
and we let $\sigma =(\sigma _{i},\sigma _{j})$ denote the strategy profile,
the set of which is still denoted $\Sigma $.

Given $\lambda \in \Lambda $ and $\sigma \in \Sigma $, we can define the
aggregate behaviors of the two players in each game, as aggregated over the
various realizations of analogy partitions.\footnote{%
We have in mind that the data available to player $i$ about $j$'s behavior
in the various games $\omega$ does not keep track of the analogy partitions
of $j$, so that the aggregate behavior constitute the only data accessible
to players.} For player $j$ in game $\omega $, this aggregate strategy can
be written as: 
\begin{equation}
\overline{\sigma }_{j}(\omega )={\displaystyle\sum_{An_{j}\in \mathcal{K}%
_{j}}}\lambda _{j}(An_{j})\sigma _{j}(\omega \mid An_{j}).  \label{agg}
\end{equation}%
Let $\bar{\sigma}=(\bar{\sigma}_{i},\bar{\sigma}_{j})$ denote a profile of
aggregate strategies and let $\overline{\Sigma }$ denote the set of such
profiles.

The analogy-based expectation of player $i$ defines for each analogy
partition $An_{i}\in $ $Supp\lambda _{i}$ and each analogy class $\alpha
_{i}\in An_{i}$, the aggregate behavior of player $j$ in $\alpha _{i}$
denoted by $\beta _{i}(\alpha _{i}\mid An_{i})\in \Delta A_{j}$ (the
dependence on $An_{i}$ is here to stress that player $i$ with analogy
partition $An_{i}$ considers only the aggregate behaviors in the various
analogy classes in $An_{i}$). Similarly as above, $\beta _{i}(\cdot \mid
An_{i})$ is said to be consistent with $\overline{\sigma }_{j}$ iff, for all 
$\alpha _{i}\in An_{i}$, 
\begin{equation}
\beta _{i}(\alpha _{i}\mid An_{i})={\displaystyle\sum_{\omega \in \alpha
_{i}}}p(\omega )\overline{\sigma }_{j}(\omega )/{\displaystyle\sum_{\omega
\in \alpha _{i}}}p(\omega )\text{.}  \label{consd}
\end{equation}%
We are now ready to propose the distributional extensions of our previous
definitions.

\begin{definition}
Given $\lambda =(\lambda _{i},\lambda _{j})\in \Lambda $, a strategy profile 
$\sigma =(\sigma _{i},\sigma _{j})\in \Sigma $ is a distributional
analogy-based expectation equilibrium (ABEE) iff there exists $\beta =(\beta
_{i},\beta _{j})$ such that for every player $i$ and $An_{i}\in Supp\lambda
_{i}$, we have that i) $\sigma _{i}(\cdot \mid An_{i})$ is a best-response
to $\beta _{i}(\cdot \mid An_{i})$ and ii) $\beta _{i}(\cdot \mid An_{i})$
is consistent with $\overline{\sigma }_{j}$ (where $\overline{\sigma }_{j}$
is derived from $\sigma _{j}$ as in (\ref{agg})).
\end{definition}

\begin{definition}
A pair $(\sigma ,\lambda )\in \Sigma \times \Lambda $ is a locally (resp.
globally) clustered distributional analogy-based expected equilibrium iff i) 
$\sigma $ is a distributional ABEE given $\lambda $, and ii) for every
player $i$ and $An_{i}\in Supp\lambda _{i}$ (where $\lambda =(\lambda
_{i},\lambda _{j})$), $An_{i}$ is locally (resp. globally) clustered with
respect to $\overline{\sigma }_{j}$ (where $\overline{\sigma }_{j}$ is
derived from $\sigma _{j}$ as in (\ref{agg})).
\end{definition}

Clearly, a clustered distributional ABEE coincides with a clustered ABEE if
the distributions of analogy partitions assign probability 1 to a single
analogy partition for both players $i$ and $j$. Clustered distributional
ABEE are thus generalizations of clustered ABEE. We now establish an
existence result.

\begin{theorem}
In finite environments, there always exists a locally (resp. globally)
clustered distributional ABEE when $d$ is the square of the Euclidean
distance or the Kullback-Leibler divergence.
\end{theorem}

We prove this result by making use of Kakutani's fixed point theorem.
Details appear in Appendix.\footnote{%
We also provide in the online Appendix B a description of the clustered
distributional ABEE in the context of the matching pennies environment of
Example 1.} An implication of clustered distributional ABEE that would
involve several analogy partitions is that different subjects exposed to the
same objective datasets (and the same memory constraints as summarized by
the number of allowed categories) may end up choosing different analogy
partitions. Such a motive for an heterogeneous way of processing the same
objective dataset is a consequence of the link between the categorizations
and the datasets (through ABEE) and it has no analog in the literature.%
\footnote{%
Our discussion about mixing is reminiscent of the observation in Nyarko
(1991) that when agents hold misspecified priors some cycling may arise.
Focusing on steady states as we do, this has led Esponda and Pouzo (2016) to
define the Berk-Nash equilibrium with possibly mixtures over models. We note
that mixing may arise in our setup because as one player changes her
clustering, she in turn affects the behavior of her opponent, and not
because the feedback received depends on the chosen action. This is
different from Nyarko's example in which cycling arises in a stationary
decision problem because the feedback received depends on the choice of the
decision maker. Another significant difference with the literature on
misspecifications is that our approach is non-parametric.}

\subsection{Learning foundation}

\label{learning}

We suggest two learning interpretations of our proposed solution concepts.
Each learning model involves populations of players randomly matched to play
the various games and two distinct stages.\footnote{%
Learning approaches consisting of two stages also appear in the literature
on learning with misspecifications, such as Fudenberg and Lanzani (2023).}
In our first learning model, subjects are exposed to raw data about
aggregate behaviors in each of the possible games in the first stage. In
this stage, they do not know yet the game that will apply to them. To make
sense of the data, they run a clustering of the raw data into $K$ clusters,
which they then use in the second stage to adjust a best-response when they
know the game they are in. In the second learning model, which has elements
of a model of cultural transmission, in the first stage, subjects inherit
categories with corresponding representative behaviors from their parents
and they best-respond to the coarse feedback in the games that apply to
them. In a second stage, the newly generated data about the plays in the
various games are observed, and these are re-arranged in new categories and
representative behaviors to be passed to the offspring. The steady states of
these learning models correspond to clustered distributional ABEE.

\noindent\textbf{Learning model 1 (Raw data and Machine Learning)}

Consider the following learning environment. There are populations of mass 1
of subjects assigned to the roles of player $1$ and $2$. At all time periods 
$t$ except the first one, there are two stages.

\begin{itemize}
\item In stage 1, subjects assigned to the role of player $i$ see the
datasets \newline
$\left\{ (\omega ,\overline{\sigma }_{j}^{t-1}(\omega ))\right\} _{\omega
\in \Omega }$, where $\overline{\sigma }_{j}^{t-1}(\omega )$ denotes the
aggregate frequencies of actions chosen at $t-1$ (and only at $t-1$) by the
subset of subjects assigned to the role of player $j$ when the game was $%
\omega $. We consider the possibility of measurement error by which we mean
that each observation $\overline{\sigma }_{j}^{t-1}(\omega )$ may be
perturbed by a (small) subject-specific idiosyncratic term. Every subject $i$
implements a clustering of his dataset into $K_{i}$ categories. That is, he
either implements a solution to (\ref{OBJ}) or he finds a local solution to
this problem (say resulting from the implementation of the K-means
algorithm). At the end of this, a subject is able to recognize to which
category/analogy class $\alpha _{i}^{t}(\omega )$ a game $\omega $ is
assigned as well as the representative point $\beta _{i}^{t}(\alpha
_{i}^{t}(\omega ))$ in $\alpha _{i}^{t}(\omega )$ defined as the mean of the
elements assigned to $\alpha _{i}^{t}(\omega )$.

\item In stage 2, subjects are randomly matched and each subject is informed
of the game $\omega $ he is in. He then expects that the subject $j$ he is
matched with will play according to $\beta _{i}^{t}(\alpha _{i}^{t}(\omega
)) $, which is the representative behavior in the analogy class to which $%
\omega $ has been assigned. Subject $i$ plays a best-response to this
expectation. At the best-response stage, we consider the possibility of
small perturbations in the payoff specifications as is commonly assumed in
learning models (Fudenberg and Kreps, 1993).
\end{itemize}

In the Online Appendix A, we establish that the dynamic model just proposed
admits steady states. Moreover, when subjects solve the full clustering
problem (i.e., solving (\ref{OBJ}) in stage 1), we show that the limit of
these steady states as the measurement error and the payoff perturbations
vanish correspond to the globally clustered distributional ABEE. These
results provide a learning foundation for the globally clustered
distributional ABEE.

It should be stressed that our learning model implicitly requires that there
is some lag between stage 1 and stage 2 so that subjects cannot in stage 2
go back to the raw data that correspond to the games they are in. In some
sense, the categorization and clustering outcome can be viewed as
representing how a constrained memory could pass information from the stage
in which the raw data are available but the relevant game is not known to
the stage in which the game is known and an action has to be chosen.%
\footnote{%
Our constraint on memory should be viewed as a physical constraint on how
many representative behaviors can be remembered (in line with Miller's
pioneering research). It should be contrasted with ideas of selective memory
that have been developed in psychology (see Fudenberg et al. (2024) for a
recent model studying the equilibrium impact of selective memory).}

\noindent\textbf{Learning model 2 (Categories, Representative Behaviors and Cultural
Transmission)}

Our second learning model involves dynasties of subjects. For each role $i$, 
each subject in the population assigned to role $i$ is viewed as
belonging to a different dynasty. At the start of period $t$ and in each
dynasty, the subject who acted at $t-1$ (the predecessor) is
replaced by a successor who receives the categories and representative
points from the predecessor. That is, each subject assigned to role $i$
receives some partition $An_{i}$ and representative points $(\beta
_{i}^{t-1}(\alpha ))_{\alpha \in An_{i}}$ from his respective predecessor. 
The share of subjects at time $t-1$ selecting $An_{i}$ in the population
assigned to role $i$ is given by $\lambda _{i}^{t-1}(An_{i})$. Then
at time $t$, the following two stages take place.

\begin{itemize}
\item In stage 1, each subject $i$ plays several games $\omega$, being
randomly re-matched each time with subjects in the role of $j$. Each time
subjects best-respond based on the inherited expectations $%
((\beta_i^{t-1}(\alpha))_{\alpha\in An_i}$.

\item In stage 2, all subjects in the role of $i$ observe the behaviors in
stage 1 of all subjects in the role of $j$ in all games $\omega $. They
compute $\bar{\sigma}_{j}^{t}(\omega )$ for each $\omega $ and they assign
games to the closest representative point $\beta _{i}^{t-1}(\alpha )$,
selecting a possibly new partition based on the local clustering criterion.
Then, a subject selecting the partition $An_{i}$ computes the representative
points $((\beta _{i}^{t}(\alpha ))_{\alpha \in An_{i}}$ to be consistent
with $\bar{\sigma}_{j}^{t}$. The share of subjects selecting partition $%
An_{i}$ in the population determines the new shares $\lambda
_{i}^{t}(An_{i})$. Each subject passes on the selected partition $An_{i}$
and the representative points to their successor within the dynasty.
\end{itemize}

While an explicit study of such a dynamic would require further work, it is
readily verified that the corresponding steady states correspond to
the locally clustered distributional ABEE.\footnote{%
Note that the clustering is made with respect to the same dataset in each
dynasty because data are aggregated over the subjects of all dynasties. This
is what ensures that steady states are locally clustered distributional ABEE
(in steady state, within a dynasty, games keep being reassigned to the same
cluster, but different dynasties may be using different clustering as long
as they are each locally clustered with respect to the stationary dataset).}%
\textsuperscript{,}\footnote{%
It should be mentioned that we would obtain the globally clustered ABEE as
steady states of a variant of learning model 2 if at the end of stage 1, we
allowed the subject to try randomly generated representative points instead
of those originally proposed by the parent, and allow the subject to pick
those if the resulting clustering led to a lower prediction error.}

\section{Applications}

In this Section we consider various applications. We focus throughout on
whether the environment has multiple self-attractive analogy partitions or
only self-repelling analogy partitions. In the former case, several choices
of analogy partitions may lead to behaviors (through the machinery of ABEE)
that agree (form the clustering perspective) to the initial choice of
analogy partitions. In the latter case, any choice of analogy partitions
leads to behaviors (through the machinery of ABEE) in at least one game that
would have to be reallocated to another analogy class (from the clustering
perspective).\footnote{%
We use the labels attractive and repelling by analogy with their use in
magnetic fields.}

While the matching penny environment discussed above gives an illustration
of an environment with self-repelling analogy partitions, we will cover more
applications in which this arises. We will also illustrate that the polar
case of self-attractive analogy partitions can arise in classic
environments, thereby suggesting a novel source of multiplicity of
equilibria in such cases.

\subsection{Self-Attractive Analogy Partitions: A Beauty Contest Game
Illustration}

We consider a family of games that induce strategic behavior in the spirit
of the \textquotedblleft beauty contest\textquotedblright\ example mentioned
in Keynes's General Theory (1936). These games are similar to those
introduced in Morris and Shin (2002) except that in our setting there is no
private information and players form their expectations in a coarse way.

Formally, a fundamental value (playing the role of the state in the above
setting) can take values $\theta \ $in $\Theta \subset 
\mathbb{R}
$ and $\theta $ is assumed to be distributed according to some smooth
density $f(\cdot )$ on $\Theta $. Player $i$ has to choose an action $%
a_{i}\in 
\mathbb{R}
$.

When the fundamental value is $\theta $, player $i$ 's utility is%
\begin{equation*}
U_{i}(a_{i},a_{-i};\theta )=-(1-r)(a_{i}-\theta )^{2}-r(a_{i}-a_{j})^{2},
\end{equation*}%
where $0<r<1$. In other words, players would like to choose an action that
is close both to the fundamental value and to the action chosen by the
opponent where $r$ measures the weight attached to the latter and $1-r$ the
weight attached to the former. It is the coordination aspect (i.e. $r>0$)
that gives to this game the flavor of the beauty contest game.\footnote{%
While these games are usually presented with more than two players, we note
that the analysis to be presented now is the same whether there are two or
more players (where in the latter case, one should require that a player
wants to coordinate with the mean action of the others). Furthermore, while
this environment has a continuum of games and a continuum of actions,
extensions of the definitions provided above for finite environments are
straightforward in this case.}

As in the framework of Section 2, players are assumed to observe $\theta $.
The quadratic loss formulation implies that if player $i$ expects $j$ to
play according to the distribution $\sigma _{j}\in \Delta 
\mathbb{R}
$ the best-response of player $i$ in game $\theta $ is

\begin{equation*}
BR(\theta ,\sigma _{j})=(1-r)\theta +rE(a_{j}\mid \sigma _{j}).
\end{equation*}

The unique Nash Equilibrium in game $\theta $ requires then that 
\begin{equation*}
a_{1}^{NE}(\theta )=a_{2}^{NE}(\theta )=\theta .
\end{equation*}%
Consider now this same environment assuming players use $K$ categories as in
Section 2. Given the symmetry of the problem, we focus on symmetric
equilibria in which players $1$ and $2$ would both choose the same analogy
partition $(\Theta _{k})_{k=1}^{K}$, and we let for each $k$%
\begin{equation*}
\overline{\theta }_{k}=E(\theta \mid \theta \in \Theta _{k})
\end{equation*}%
denote the mean of the fundamental values conditional on $\theta $ lying in
the analogy class $\Theta _{k}$. Straightforward calculations (detailed for
completeness in the online Appendix B) show that there is a unique
analogy-based expectation equilibrium such that for $\theta \in \Theta _{k}$%
: 
\begin{equation}
a_{1}^{ABEE}(\theta )=a_{2}^{ABEE}(\theta )=(1-r)\theta +r\overline{\theta }%
_{k}.  \label{ABEE}
\end{equation}%
In this equilibrium, players do not choose the fundamental value $\theta $
because their coarse expectation leads them to expect the mean action of the
opponent to be $\overline{\theta }_{k}$ and not $\theta $ when $\theta \in
\Theta _{k}$.

Assume that players use the square of the Euclidean distance applied to the
mean of the distribution for clustering purposes. That is, the prediction
error measure between the behavior of player $j$ in $\theta $ and the
aggregate behavior of player $i$ in the the subset $\Theta _{k}^{\prime }$
is given by $(a_{j}^{ABEE}(\theta )-E(a_{j}^{ABEE}(\theta ^{\prime })\mid
\theta ^{\prime }\in \Theta _{k}^{\prime }))^{2}$.


Our main insight is the observation that there are many possible clustered
analogy-based expectation equilibria when the concern for coordination is
large enough. More precisely,

\begin{proposition}
Take any partition $(\Theta _{k})_{k=1}^{K}$ such that $\overline{\theta }%
_{k}=E(\theta \mid \theta \in \Theta _{k})$ are all different. Then for $r$
sufficiently close to $1$, when both players use this analogy partition and
play according to (\ref{ABEE}), we have a Clustered Analogy-based
Expectation Equilibrium.
\end{proposition}

\textbf{Proof}. When $r$ is close to 1, actions $a_{j}^{ABEE}(\theta )$ in $%
\Theta _{k}$ are all close to $\overline{\theta }_{k}$. When $\overline{%
\theta }_{k}$ are all distinct, the clustering (whether local or global)
leads to $(\Theta _{k})_{k=1}^{K}$. \textbf{Q.E.D.}

In other words, our beauty contest game illustrates in a stark way the
possibility of self-attractive analogy partitions. When $r$ is close to $1$,
virtually all analogy partitions can be sustained as part of a clustered
ABEE. Or to put it differently: Once the analogy partition $(\Theta
_{k})_{k=1}^{K}$ is chosen and no matter what this partition is, players are
led through the working of the ABEE to behave in a way that makes the
clustering into $(\Theta _{k})_{k=1}^{K}$ best for the purpose of minimizing
prediction errors. Observe that the vast multiplicity of analogy partitions
so derived results in a vast range of possible equilibrium behaviors as
well, where behaviors are concentrated around the various $\bar{\theta}_k$.%
\footnote{%
Observe also that our construction would be robust to the introduction of
any share of rational agents to the extent that in the limit as $r$ tends to
1 players in our equilibrium are picking best-responses to their opponent's
strategy.}

Of course, we should not expect to have such an extreme form of
self-attraction for all parameter values of the beauty contest game. For
example, when $r$ is small (close to $0$), then behaviors are not much
affected by the analogy partition, and in the limit as $r=0$, the only
globally clustered ABEE would require choosing $(\Theta _{k})_{k=1}^{K}$ so
as to minimize 
\begin{equation*}
\sum_{k=1}^{K}\int_{\Theta _{k}}(\theta -\overline{\theta }_{k})^{2}f(\theta
)d\theta .
\end{equation*}%
In the case $\theta $ is uniformly distributed on $\left[ \underline{\theta }%
,\overline{\theta }\right] $, this would lead to the equal splitting analogy
partition (i.e., $\Theta _{k}=(\theta _{k-1},\theta _{k})$ where $\theta
_{0}=\underline{\theta }$ and $\theta _{k}=\underline{\theta }+\frac{k}{K}(%
\overline{\theta }-\underline{\theta })$).

For intermediate values of $r$, we can support a bigger range of analogy
partitions as part of a CABEE but not as many as when $r$ approaches $1$.
The next Proposition establishes that as $r$ grows larger, more and more
analogy partitions can be obtained.\footnote{%
The same result would hold for globally clustered ABEE (even if a detailed
proof is a bit tedious to derive). A weaker statement that can easily be
established is that if a partition is part of a globally CABEE at $r$, then
there exists $r^{\ast }$ such that it is also part of a globally CABEE at
all $r^{\prime }\geq r^{\ast }$.}

\begin{proposition}
Take a partition $(\Theta _{k})_{k=1}^{K}$ and the corresponding ABEE for
some $r<1$. Suppose this is a locally CABEE. Then, for any $r^{\prime }>r$, $%
(\Theta _{k})_{k=1}^{K}$ and the corresponding ABEE form a locally CABEE.
\end{proposition}

\textbf{Proof.} For each player $i$ note the following observations for a partition $(\Theta _{k})_{k=1}^{K}$. As $r$
increases, for each $\theta \in \Theta _{k}$, $a_j^{ABEE}(\theta )$ gets
(weakly) closer to $\bar{\theta}_{k}$. Moreover if $\theta \in \Theta _{k}$
is weakly greater (resp. smaller) that $\bar{\theta}_{k}$, then: 

(i) for each $\bar{\theta}_{k^{\prime }}>\bar{\theta}_{k}$ (resp. $%
<$), if the partition is locally clustered with respect to the
corresponding ABEE for some $r<1$, then $a_{j}^{ABEE}(\theta )$ for $r'>r$ is strictly further apart from $\bar{\theta}_{k^{\prime }}$ as compared to $a_{j}^{ABEE}(\theta )$ for $r$;\footnote{This holds because if $a_{j}(\theta )\geq\bar{\theta}_{k^{\prime }}$ for
some $\theta \in \Theta _{k}$ and some $\bar{\theta}_{k'}>\bar{\theta}%
_{k}$, then the partition is not locally clustered with respect to 
$a_{j}(\cdot )$.}

(ii) for each $\bar{\theta}_{k^{\prime }}\leq\bar{\theta}_k$
(resp. $\geq$), $a_j^{ABEE}(\theta)$ is weakly closer to $\bar{\theta}_k$
than $\bar{\theta}_{k^{\prime }}$ because $a_j^{ABEE}(\theta)$ is weakly
greater (resp. smaller) than $\bar{\theta}_k$ for any $r$.



Thus, if $(\Theta _{k})_{k=1}^{K}$ is locally clustered with respect to $%
a_j^{ABEE}(\cdot)$ for given $r$, then it is also locally clustered with
respect to $%
a_j^{ABEE}(\cdot)$ for any $r^{\prime }>r$. \textbf{Q.E.D.}

The insight of this Proposition can possibly be related to some features of
echo chambers. When the concern for coordination is big, the clustering of
fundamental values into analogy classes may not be much related to the
objective realization of the fundamental values, and different societies
(which may settle down on different CABEE) may end up forming different
beliefs and adopting different behaviors in objectively identical situations.

The analysis presented here should be contrasted with insights obtained in
the tradition of global games, suggesting a unique selection of equilibrium
(see Morris and Shin (2002) for references). Of course, our setting is
different (no private information), and our formalization of expectations
through categories is also different, leading to alternative predictions and
a novel perspective on the possibility of multiple equilibria. We believe
that our finding that a larger variety of behaviors and larger departures
from the fundamental can be obtained as agents are more concerned about
coordinating with others agrees with Keynes' intuition about beauty
contests. While rational expectations in the beauty contest game would lead
agents to adopt behaviors coinciding with the fundamental value, in our
approach a much wider variety of behaviors and expectations can arise in
such cases.

\subsection{Self-Repelling Analogy Partitions: A Monitoring Game Illustration%
}

Unless specified otherwise, the results stated in this subsection hold for $%
d $ being either the square of the Euclidean distance or the
Kullback-Leibler divergence. We consider the following Employer-Worker
environment. There are three types of workers $a,b,c$. An employer is
matched with one worker who can be either of type $a$, $b$, or $c$ with
probability $p_{a},p_{b},p_{c}$, respectively. The type $\omega =a,b,c$ of
the worker is observed by the employer. It can be identified with the state
in our general framework.

In each possible interaction $\omega $, the employer and the worker make
simultaneous decisions. The employer decides whether he will Control ($C$)
the worker or not ($D$). The worker decides whether to exert low effort $e=0$
or high effort $e=1$.

We assume that the employer cares about the type only through the effort
attitude that she attributes to the type of worker. Specifically, we assume
that the employer finds it best to choose $C$ only if she expects the worker
to choose low effort with probability no less than $\nu ^{\ast }$.

Workers' effort attitudes depend on their type and possibly on their
expectation about whether they will be controlled or not. Type $a$ of worker
always chooses $e=0$ (irrespective of his expectation about the Control
probability). Type $b$ always chooses $e=1$. Type $c$ of worker finds it
best to exert low effort $e=0$ when he expects $C$ to be chosen with
probability no more than $\mu ^{\ast }$.

In the unique Nash equilibrium, the employer would choose $C$ when facing
type $a$, $D$ when facing type $b$ and would mix between $C$ with
probability $\mu ^{\ast }$ and $D$ with probability $1-\mu ^{\ast }$ when
facing type $c$. Type $c$ of worker would choose $e=0$ with probability $\nu
^{\ast }$.

Assuming the employer uses two categories $K=2$ whereas the worker is
rational,\footnote{%
One can possibly motivate this asymmetry by saying that each type of worker
can safely focus on the situations in which similar types are involved
whereas employers need to have a more complete understanding on how the type
maps into an effort attitude.} we have:

\begin{proposition}
Assume $p_{c}$ is no larger than $p_{a}$ and $p_{b}$, $\frac{p_{c}}{%
p_{a}+p_{c}}<\nu ^{\ast }<\frac{p_{b}}{p_{b}+p_{c}}$, and $\nu ^{\ast }\neq 
\frac{p_{b}}{p_{a}+p_{b}}$. There is no pure Clustered Analogy-based
expectation equilibrium in which a single analogy partition is used by the
employer.
\end{proposition}

\textbf{Proof}. This is proven by contradiction. Let $\beta _{e=1}$ denote
the probability attached to $e=1$ by the employer in her non-singleton
analogy class. 1) If $ac$ are put together, consistency implies that $\beta
_{e=1}$ is at most $\frac{p_{c}}{p_{a}+p_{c}}$, which is smaller than $\nu
^{\ast }$. Best-response of the employer implies that $C$ (Control) is
chosen when facing worker $c$. Best-response of the $c$-worker leads to $%
e_{c}=1$. The resulting profile of effort attitude is $%
(e_{a}=0,e_{b}=1,e_{c}=1)$, which leads at the clustering stage to reassign $%
c$ with $b$, thereby yielding a contradiction. 2) If $bc$ are put together, $%
\beta _{e=1}>\nu ^{\ast }$ so that $D$ is chosen by the employer when facing
the $c$-worker. This leads the $c$-worker to choose $e_{c}=0$ and the
resulting profile of effort attitudes $(e_{a}=0,e_{b}=1,e_{c}=0)$ would lead
to reassign $c$ to $a$, thereby leading to a contradiction. 3) If $ab$ are
put together, $\beta _{e=1}=\frac{p_{b}}{p_{a}+p_{b}}$, and the probability
that $e_{c}=1$ is $\nu ^{\ast }$ (as the mixed Nash equilibrium is then
played in the interaction with $c$). This violates the condition for local
clustering, for either $a$ or $b$ whenever $\nu ^{\ast }\neq \frac{p_{b}}{%
p_{a}+p_{b}}$. \textbf{Q.E.D.}

In other words, in this monitoring environment, any categorization is
self-repelling, and steady state requires some mixing. When considering
global clustering, we have:

\begin{proposition}
Assume $p_{c}$ is no larger than $p_{a}$ and $p_{b}$ and $\frac{p_{c}}{%
p_{a}+p_{c}}<\nu ^{\ast }<\frac{p_{b}}{p_{b}+p_{c}}$. There is a unique
globally Clustered Distributional ABEE in which the analogy partition
putting $ac$ (resp. $bc$) together is chosen with probability $\mu ^{\ast }$
(resp. $1-\mu ^{\ast }$), the $c$ worker chooses $e=0$ and $1$ each with
probability half, and the employer chooses $C$ in any analogy class
containing $a$ and $D$ in any analogy class containing $b$.
\end{proposition}

This can easily be understood as follows. Given the choice of the Employer
for each of her analogy partition, the distribution over analogy partitions
is chosen so as to make the $c$ worker indifferent between his two effort
options. The mixing of the $c$ worker is then chosen so as to make the two
analogy partitions equally good for the purpose of minimizing the total
variance (when $d$ is the square of the Euclidean distance) or the relative
entropy (when $d$ is the Kullback-Leibler divergence). The rest of the
construction follows easily.

One noteworthy aspect of the equilibrium is that $\nu ^{\ast }$ plays no
role in the equilibrium (so long as $\frac{p_{c}}{p_{a}+p_{c}}<\nu ^{\ast }<%
\frac{p_{b}}{p_{b}+p_{c}}$). This is so because, in equilibrium, the
employer is made indifferent between the two possible partitions at the
clustering stage, but then her monitoring choice is deterministic for each
analogy partition. Instead, Nash equilibrium prescribes that the employer is
made indifferent between her monitoring options, thereby implying that the
worker should exert effort $e=1$ with probability $\nu ^{\ast }$.

Similar insights are obtained when local clustering is considered: a wider
range of equilibria can be sustained, but in all of them, the analogy
partition putting $ac$ (resp. $bc$) together must be chosen with probability 
$\mu ^{\ast }$ (resp. $1-\mu ^{\ast }$). The range of these depend on the
probabilities $p_{\omega } $ of $\omega =a,b,c$ and to fix ideas, we
consider in the next proposition that types $a$ and $b$ are equally likely.

\begin{proposition}
Assume that $p_{a}=p_{b}=p>\frac{1}{3}$, $\frac{p_{c}}{p_{a}+p_{c}}<\nu
^{\ast }<\frac{p_{b}}{p_{b}+p_{c}}$, and $\nu^*\neq\frac{1}{2}$. The
following defines the locally Clustered Distributional ABEE. The analogy
partition putting $ac$ (resp. $bc $) together is chosen with probability $%
\mu ^{\ast }$ (resp. $1-\mu ^{\ast }$). The employer chooses $C$ in any
analogy class containing $a$ and $D$ in any analogy class containing $b$.
The $c$ worker chooses $e=0$ with probability $\zeta $ in the range $(p, 
\frac{1-p}{2-3p})$ when $d$ is the squared Euclidean distance and in the
range $(0,1)$ when $d$ is the Kullback-Leibler divergence.
\end{proposition}

The main difference between the above two Propositions is that for local
clustering, any mixing of the $c$ type that assigns probability $\zeta \in
(p,\frac{1-p}{2-3p})$ on $e=0$ makes the two analogy partitions locally
clustered with respect to the proposed strategy of the worker. Indeed, when
the analogy partition putting $ac$ (resp. $bc$) together is considered, the
aggregate effort distribution in $ac$ (resp. $bc$) assigns probability $%
\frac{p+(1-2p)\zeta }{1-p}$ (resp. $\frac{(1-2p)}{1-p}\zeta $) to $e=0$.
When $d$ is the square of the Euclidean distance, the mixing $\zeta $ on $%
e=0 $ is then closer to $\frac{p+(1-2p)\zeta }{1-p}$ (resp. $\frac{(1-2p)}{%
1-p}\zeta $) than to $0$ (resp. $1$) since $\zeta >p$ (resp. $\zeta <\frac{%
1-p}{2-3p}$). The Kullback-Leibler divergence is more permissive as it
explodes to infinity when the support of the behavior is not contained in
the support of the expectations. Then, any $\zeta \in (0,1)$ can be
sustained because the behavior of $c$ worker assigns positive probability to
both $e=1$ and $e=0$, while the expectations of the singleton classes only
to one of the two.

With the above learning interpretation that involves populations of
employers and workers, the observation that several analogy partitions must
arise in a clustered distributional ABEE imply that not all employers
categorize $c$ workers in the same way. This implies that different
employers may have different beliefs about the working attitude of $c$
workers. Some employers (in proportion $\mu ^{\ast }$) believe $c$ workers
choose $e=0$ with probability $\frac{p+(1-2p)\zeta }{1-p}$ and others (in
proportion $1-\mu ^{\ast }$) believe $c$ workers choose $e=0$ with
probability $\frac{(1-2p)}{1-p}\zeta $. That is, some overestimate the
working attitude of $c$ workers and some underestimate it, leading in our
model to polarized beliefs. Those employers underestimating the working
attitude will treat the minority group ($c$ is the least likely type)
exactly like type $a$, while the others will treat them exactly like type $b$%
. Even if stylized, we believe our analysis may provide a novel argument as
to how different beliefs about the working attitude of minority groups and
different treatments of said groups may co-exist in a society.

\subsection{Strategic interactions with linear best-replies}

In this section we consider families of games with continuous action spaces
parametrized by an interaction parameter $\mu $, which takes values in an
interval of the real line. This parameter is a determinant of the intensity
of players' reactions to their opponent's behavior. Players have
best-responses which are linear both in the strategy of the opponent and in $%
\mu $.

Formally, we consider a family of games parametrized by $\mu \in \lbrack
-1,1]$, where $\mu $ is distributed according to a continuous density
function $f$ with cumulative denoted by $F$. Players observe the realization
of $\mu $ and player $i=1,2$ chooses action $a_{i}\in \mathbb{R}$.\footnote{%
The environment we consider here has a continuum of games and a continuum of
actions. While our general existence results do not apply to such
environments, we will obtain the existence of locally clustered ABEE in the
case of strategic complements by construction. The case of strategic
substitutes will be shown not to admit locally clustered ABEE.} In game $\mu 
$, when player $i$ expects player $j$ to play according to $\sigma _{j}\in
\Delta \mathbb{R}$, player $i$'s best-response is: 
\begin{equation*}
BR_{i}(\mu ,\sigma _{j})=A+\mu B+\mu CE_{\sigma _{j}}(a_{j}),
\end{equation*}%
where $A,B$\ and $C$ are constants with $0<C<1$ and $\sigma _{j}$ is
required to be such that the resulting mean action $E_{\sigma _{j}}(a_{j})$
is well defined.\footnote{%
We will check that this automatically holds for best-responses in our
construction.} We will analyze separately the cases in which $\mu \in
\lbrack 0,1]$\ and $\mu \in \lbrack -1,0]$, and in each case we will assume
that $f(\cdot )$\ has full support. In the former case, the games exhibit
strategic complementarity. In the latter, they exhibit strategic
substitutability.

The restriction to linear best-replies while demanding allows us to
accommodate classic applications.\footnote{%
It may be mentioned that in our formulation, we allow the actions to take
any value (positive or negative) whereas in some of the applications
mentioned below it would be natural to impose that the actions (quantities,
prices or effort level) be non-negative. We do not impose non-negativity
constraints to avoid dealing with corner solutions, but none of our
qualitative insights would be affected by such additional constraints.}
Consider duopolies with differentiated products, constant marginal costs and
linear demands. Whether firms compete in prices \`{a} la Bertrand or in
quantities \`{a} la Cournot, best-responses are linear, and we have
strategic complements in the former case and strategic substitutes in the
latter (see Vives (1999) for a textbook exposition). Considering games with
different values of $\mu $ can be interpreted as considering different
duopoly environments with different market/demand conditions for such
duopolies.

Regardless of the sign of $\mu $, it is readily verified that there exists a
unique Nash Equilibrium of the game with parameter $\mu $. It is symmetric,
it employs pure strategies and it is characterized by $a_{1}^{NE}(\mu
)=a_{2}^{NE}(\mu )=\frac{A+\mu B}{1-\mu C}$.

We will be considering analogy partitions with the property that each
analogy class is an interval of $\mu $, and we will refer to these as
interval analogy partitions. Specifically, assume that players use (pure)
symmetric interval analogy partitions, splitting the interval into $K$
subintervals, so that 
\begin{equation*}
An_{1}=An_{2}=\{[\mu _{0},\mu _{1}],(\mu _{1},\mu _{2}],\dots ,(\mu
_{K-1},\mu _{K}]\}
\end{equation*}%
where $\mu _{0}=0,\,\mu _{K}=1$ in the case of strategic complements, and $%
\mu _{0}=-1,\mu _{K}=0$ in the case of strategic substitutes.\footnote{%
Whether $\mu _{k}$ is assigned to $(\mu _{k-1},\mu _{k})$ or $(\mu _{k},\mu
_{k+1})$ plays no role in our setting with a continuum of $\mu $.}

Since analogy partitions are symmetric, we simplify notation by dropping the
subscript that indicates whether player $1$ or $2$ is considered. We simply
denote the interval $(\mu _{k-1},\mu _{k}]$ by $\alpha _{k}$ for $k=2,...K-1$
and $\alpha _{1}=[\mu _{0},\mu _{1}]$.

Given that players care only about the mean action of their opponent, we
will assume that players focus only on this mean and accordingly compare the
behaviors in different games using the squared Euclidean distance between
the mean action these games induce. That is, we will consider the square of
the Euclidean distance in the space of these mean actions for clustering
purposes.

Specifically and with some abuse of notation, we will refer to $\beta
_{i}(\alpha _{k})$ as the expected mean action of player $j$ in the analogy
class $\alpha _{k}$. The consistency of $\beta _{i}$ with $\sigma _{j}$
imposes that $\beta _{i}(\alpha _{k})=\frac{1}{F(\mu _{k})-F(\mu _{k-1})}%
\int_{\mu _{k-1}}^{\mu _{k}}\sigma _{j}(\mu )f(\mu )d\mu $, where $\sigma
_{j}(\mu)$ denotes the (mean) action chosen by player $j$ in game $\mu $.

Given a (symmetric) interval analogy partition profile $An_{1}=An_{2}$, an
ABEE is a strategy profile $(\sigma _{1},\sigma _{2})$ such that for each
player $i$, each class $\alpha _{k}$ and each game $\mu \in \alpha _{k}$, we
have $\sigma _{i}(\mu )\in BR_{i}(\mu ,\beta _{i}(\alpha _{k}))$ with the
requirement that $\beta _{i}$ is consistent with $\sigma _{j}$. It is
readily verified (through routine calculations provided in the Online
Appendix B) that there exists a unique ABEE, which is symmetric.

\begin{proposition}
\label{LINABEE}Assume players use symmetric interval analogy partitions.
There exists a unique ABEE where, for all $k=1,\dots ,K$, $\beta _{1}(\alpha
_{k})=\beta _{2}(\alpha _{k})=\frac{A+BE[\mu |\alpha _{k}]}{1-CE[\mu |\alpha
_{k}]}$ and for $\mu \in \alpha _{k}$, $\sigma _{1}(\mu )=\sigma _{2}(\mu
)=A+\mu \frac{B+AC}{1-CE[\mu |\alpha _{k}]}$.
\end{proposition}

Since under symmetric interval analogy partitions the ABEE is symmetric, we
drop the subscript that refers to players and we write $\beta _{1}(\alpha
_{k})=\beta _{2}(\alpha _{k})=\beta (\alpha _{k})$. We also let $a(\mu
|\alpha _{k})$ refer to $A+\mu \frac{B+AC}{1-CE[\mu |\alpha _{k}]}$ for the
remainder of this section. Local clustering would require that in each game $%
\mu \in \alpha _{k}$, the action $a(\mu |\alpha _{k})$ be (weakly) closer to 
$\beta (\alpha _{k})$ than to any alternative $\beta (\alpha _{k^{\prime }})$
for any $k^{\prime }\neq k$.

Global clustering on the other hand would require that the given candidate
interval analogy partition $An=\{\alpha _{k}\}_{k=1}^{K}$\ solves 
\begin{equation*}
An=\arg \min_{\{\alpha _{k}^{\prime }\}_{k=1}^{K}}\sum_{k}\int_{\alpha
_{k}^{\prime }}\left[ \beta (\alpha _{k}^{\prime })-a^{ABEE}(\mu )\right]
^{2}f(\mu )d\mu
\end{equation*}%
where $a^{ABEE}(\mu )\equiv A+\mu \sum_{k=1}^{K}1_{\{\mu _{k-1}<\mu \leq \mu
_{k}]\}}\frac{B+AC}{1-CE[\mu |\alpha _{k}]}$ is the ABEE strategy given $An$
and $\beta (\alpha _{k}^{\prime })=E(a^{ABEE}(\mu )\mid \mu \in \alpha
_{k}^{\prime })$.

\subsubsection{Strategic Complements}

In this part, we consider the strategic complements case, and we assume that 
$\mu $ is distributed according to a continuous density $f$ with support on $%
[0,1]$.

Given $\left( \mu _{k}\right) _{k=0}^{K}$, we wish to highlight that 
\begin{equation*}
a^{ABEE}(\mu )=A+\mu \sum_{k=1}^{K}\mathbf{1}_{\{\mu _{k-1}<\mu \leq \mu
_{k}]\}}\frac{B+AC}{1-CE[\mu |\alpha _{k}]}
\end{equation*}%
has discontinuities at $\mu _{1},\mu _{2},\dots ,\mu _{K-1}$. If $B\geq -AC$%
, the function $a^{ABEE}(\mu )$ is increasing in $\mu $ and the
discontinuities take the form of upward jumps. Similarly, if $B<-AC$, the
function $a^{ABEE}(\mu )$ is decreasing in $\mu $ and the discontinuities
take the form of downward jumps. The direction of the jumps is a consequence
of the strategic complement aspect, and it will play a key role in the
analysis of local clustering. Indeed assuming $B\geq -AC$, as one moves in
the neighborhood of $\mu _{k}$ from the analogy class $(\mu _{k-1},\mu _{k}]$
to the analogy class $(\mu _{k},\mu _{k+1}]$, the perceived mean action of
the opponent jumps upwards and this leads to an upward jump in the
best-response due to strategic complementarity.

There is a simple geometric characterization of local clustering. Assuming $%
B\geq -AC$,\footnote{%
The case in which $B<-AC$ is analyzed symmetrically.} we have that $\beta
(\alpha _{k})\leq \beta (\alpha _{k+1})$, for all $k$. The local clustering
requirements are equivalent to the condition that the arithmetic average of
the analogy-based expectations of two adjacent analogy classes should be
between the largest action in the first and the smallest action in the
second analogy class. That is, 
\begin{equation*}
a(\mu _{k}|\alpha _{k})\leq \frac{\beta (\alpha _{k})+\beta (\alpha _{k+1})}{%
2}\leq a(\mu _{k}|\alpha _{k+1}),
\end{equation*}%
which can be assessed graphically by plotting the functions $a(\mu |\alpha
_{k})$ (see Figure 1, (a)).\footnote{%
Figure 1 shows how the (simplified) local clustering requirements would
appear graphically. There are two graphs, one for $B>-AC$ on the left and
one for $B<-AC$ on the right. Both graphs depict how the Nash Equilibrium
function $a^{NE}(\mu )=\frac{A+\mu B}{1-\mu C}$ (in blue) and the ABEE
function $a^{ABEE}(\mu )$ (in orange) change as $\mu $ varies. For these
graphs we assume that $\mu $ is distributed uniformly over $[0,1]$, we let $%
K=4$, and we pick the interval analogy partition induced by the equal
splitting sequence $\{0,\frac{1}{4},\frac{2}{4},\frac{3}{4},1\}$. When $%
B>(<)AC$, $a^{NE}(\mu )$ and $a^{ABEE}(\mu )$ are strictly increasing
(decreasing).}

\begin{figure}[h!]
\begin{adjustwidth}{1.5mm}{1.5mm}
\caption{\begin{small}Local Clustering for $K=4$ and $\protect\mu$-sequence such that $%
\protect\mu_k=\frac{k}{4}$, $k=0,1,\dots,4$.\end{small}}%
\begin{subfigure}{0.48\textwidth}
    \includegraphics[width=\textwidth]{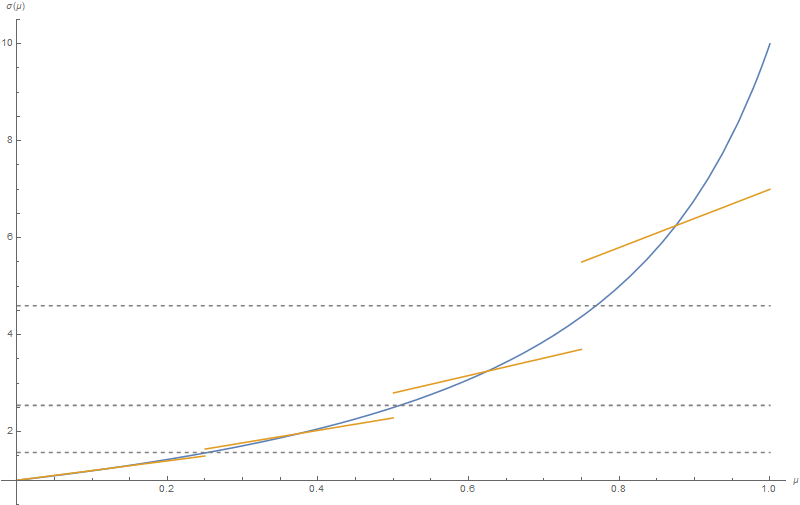}
    \caption{Increasing NE and ABEE functions\\ 
		with $A=1$,$B=1$ and $C=0.8$}
  \end{subfigure}
\hfill 
\begin{subfigure}{0.48\textwidth}
    \includegraphics[width=\textwidth]{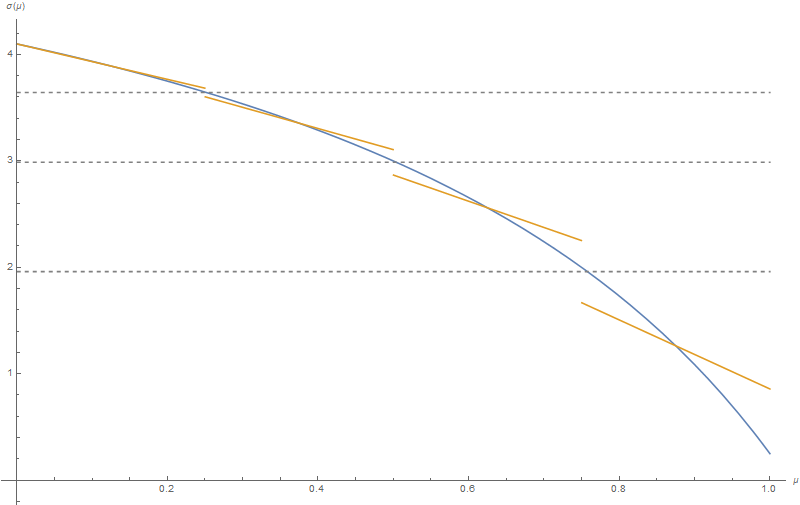}
    \caption{Decreasing NE and ABEE functions\\ 
		with $A=4.1$,$B=-4$ and $C=0.6$}
  \end{subfigure}
\end{adjustwidth}
\end{figure}

To present our result, we introduce the notion of equidistant-expectations
sequence $\mu _{0},\mu _{1},\dots ,\mu _{K}$ defined so that for any $\mu
_{k}$, with $k\neq 0,1$, the Euclidean distance between $\mu _{k}$ and the
mean value of $\mu $ in $(\mu _{k-1},\mu _{k}]$ is equal to the Euclidean
distance between $\mu _{k}$ and the mean value of $\mu $ in $(\mu _{k},\mu
_{k+1}]$. That is, $\mu _{k}-E[\mu |(\mu _{k-1},\mu _{k}]]=E[\mu |(\mu
_{k},\mu _{k+1}]]-\mu _{k}$. We refer to the corresponding interval
partition $(\alpha _{k})_{k=1}^{K}$ with $\alpha _{k}=(\mu _{k-1},\mu _{k}]$
as the equidistant-expectations partition. We note that when $\mu $\ is
uniformly distributed on $[0,1]$, the equidistant-expectations sequence is
uniquely defined by $\mu _{k}=\frac{k}{K}$. For more general density
functions $f$, it is readily verified (by repeated application of the
intermediate value theorem) that there always exists at least one
equidistant-expectations partition.\footnote{%
See the Lemma \ref{equid} in the Online Appendix B for details.}

Our main result in the strategic complements case is:

\begin{proposition}
\label{LOCALCOMP}In the environment with strategic complements, consider an
equidistant-expectations partition $\left( \mu _{k}^{\ast }\right)
_{k=0}^{K} $. There exist $\left\{ \underline{\mu }_{k},\overline{\mu }%
_{k}\right\} _{k=0}^{K}$satisfying $\underline{\mu }_{k}<\mu _{k}^{\ast }<%
\overline{\mu }_{k}$ such that for any $\left( \mu _{k}\right) _{k=0}^{K}$
satisfying $\mu _{k}\in (\underline{\mu }_{k},\overline{\mu }_{k})$, the
analogy partition $An=(\alpha _{k})_{k=1}^{K}$ with $\alpha _{k}=(\mu
_{k-1},\mu _{k}] $ together with the corresponding ABEE is a locally
clustered ABEE.
\end{proposition}

The rough intuition for this result can be understood as follows. First,
start with an equidistant-expectations partition $\left( \mu _{k}^{\ast
}\right) _{k=0}^{K}$. Suppose we were considering games with no interaction
term, i.e., such that $C=0$. Then in game $\mu $, players would be picking
their dominant strategy $a(\mu )=A+\mu B $ irrespective of the analogy
partition, given that players would not care about the action chosen by
their opponent. It is readily verified that the local clustering conditions
for the points $a(\mu )$ would lead to pick an equidistant-expectations
partition in this case. Allowing for non-null interaction parameters $C$
makes the problem of finding a locally clustered ABEE a priori non-trivial
due to the endogeneity of the data generated by the ABEE with respect to the
chosen analogy classes, as emphasized above. However, what the Proposition
implies is that using the same analogy classes as those obtained when $C=0$
can be done to construct a locally clustered ABEE. Intuitively, this is so
because the strategic complement aspect makes the points obtained through
ABEE in a given class of the equidistant-expectations partition look closer
to one another relative to points outside a class, as compared with the case
in which $C=0$. As a result, the local clustering conditions which hold for
the equidistant-expectation partition when $C=0$ hold a fortiori when $C$ is
non-null. Now given the jumps, there is some slack in the conditions for
local clustering, thereby ensuring that one can find open intervals for the
boundary points $\mu _{k}$ that satisfy the conditions of the Proposition.

Our environment with strategic complements can be viewed as one in which
there is an element of self-attraction in the choice of analogy partitions.
As can easily be inferred from the above discussion, the larger $C$ is, the
more analogy interval analogy partitions can be used to support locally
clustered ABEE, i.e., the more partitions are self-attractive. In Jehiel and
Weber (2023) (the working paper version), we have characterized more fully
the set of locally clustered ABEE confirming that insight, and we have noted
that when $C$ is too large the equidistant-expectations partition need not
be compatible with the requirement for global clustering.\footnote{%
While our analysis there suggests that intervals with larger $\mu $ should
be smaller than in the equidistant-expectations partition, a general
characterization of globally clustered ABEE in this case should be the
subject of further research.}

\subsubsection{Strategic Substitutes}

We consider now the strategic substitute case, and we assume that $\mu $ is
distributed according to a continuous density function $f$ with support $%
[-1,0]$.

Note that, differently from the strategic complements environment, here the
ABEE function $a^{ABEE}(\mu )=A+\sum_{k=1}^{K}1_{\{\mu \in \lbrack \mu
_{k-1},\mu _{k})\}}\mu \frac{B+AC}{1-CE[\mu |\alpha _{k}]}$ is not monotonic
in $\mu $. This is due to the fact that at the discontinuity points the
jumps of the function are in the opposite direction with respect to the
slope of $a(\mu |\alpha _{k})=A+\mu \frac{B+AC}{1-CE[\mu |\alpha _{k}]}$,
and this is a fundamental difference induced by the change from strategic
complements to strategic substitutes. This is illustrated in Figure 2.

\begin{figure}[!h]
\begin{adjustwidth}{1.5mm}{1.5mm}
\caption{\begin{small}Local Clustering for $K=4$ and $\protect\mu$-sequence such that $%
\protect\mu_k=\frac{k}{4}$, $k=0,1,\dots,4$.\end{small}}%
\begin{subfigure}{0.48\textwidth}
    \includegraphics[width=\textwidth]{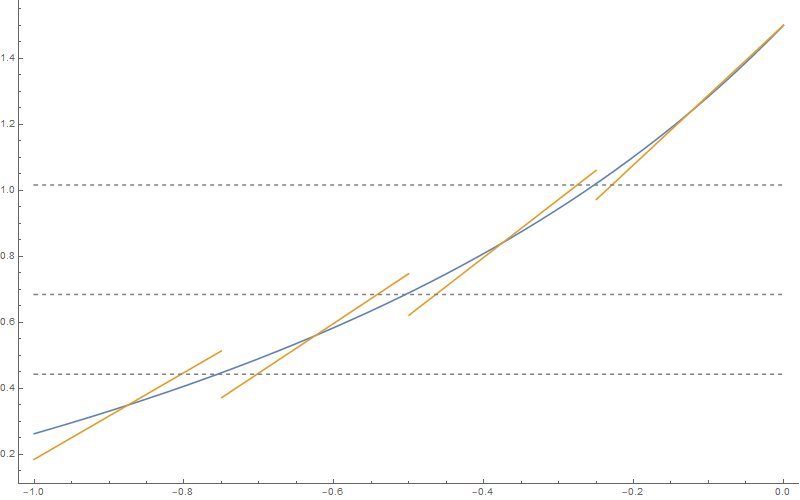}
    \caption{Increasing NE and ABEE functions\\
		with $A=1.5$,$B=1$ and $C=0.9$}
  \end{subfigure}
\hfill 
\begin{subfigure}{0.48\textwidth}
    \includegraphics[width=\textwidth]{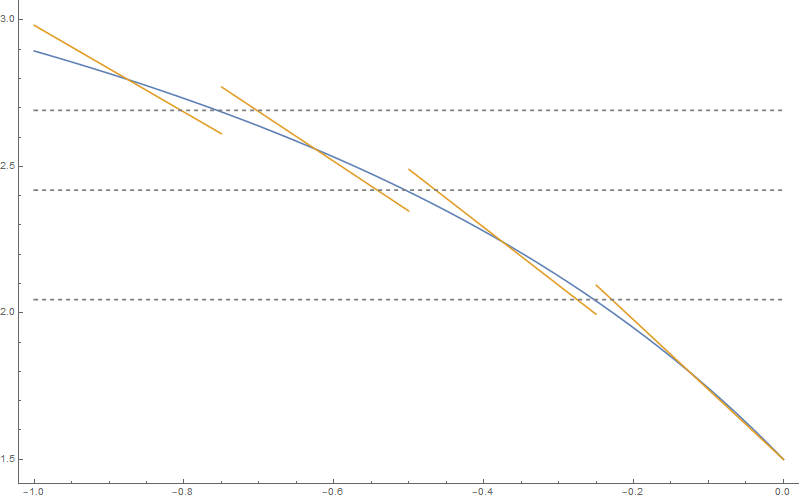}
    \caption{Decreasing NE and ABEE functions\\
		with $A=1.5$,$B=-4$ and $C=0.9$}
  \end{subfigure}
\end{adjustwidth}
\end{figure}

The non-monotonicities that arise in the ABEE strategies with interval
analogy partitions in turn make it impossible to satisfy the local
clustering conditions in a neighborhood of $\mu =\mu _{k}$. This is so
because it cannot be simultaneously the case that $\lim_{\mu \rightarrow \mu
_{k}^{+}}a^{ABEE}(\mu )$ is closer to $\beta ((\mu _{k},\mu _{k+1}])$ and $%
a^{ABEE}(\mu _{k})$ is closer to $\beta ((\mu _{k-1},\mu _{k}])$.\footnote{%
For example when $B+AC>0$, we would have $\beta ((\mu _{k},\mu
_{k+1}])>a^{ABEE}(\mu _{k})>\lim_{\mu \rightarrow \mu _{k}^{+}}a^{ABEE}(\mu
_{k}^{+})>\beta ((\mu _{k-1},\mu _{k}])$, making it impossible to satisfy
the local clustering conditions for $\mu =\mu _{k}$ and $\mu =\mu _{k}^{+}$
(i.e., $\mu $ slightly above $\mu _{k}$).} This simple observation implies:

\begin{proposition}
\label{SUB}In the strategic substitutes environment, whenever $B\neq -AC$,
there are no symmetric interval analogy partitions that are locally
clustered with respect to the induced ABEE.
\end{proposition}

Our environment with strategic substitutes illustrates a case with
self-repelling analogy partitions. In light of our general theoretical
framework, it would then be natural to look for clustered distributional
ABEE. In our setting with a continuum of games, this is a challenging task,
and in Jehiel and Weber (2023) (the working paper version), we characterize
the clustered distributional ABEE when $\mu $ can take only three values.

\section{Conclusion}

In this paper we have introduced the notion of Clustered ABEE defined so
that i) given the analogy partitions, players choose strategies following
the ABEE machinery, and ii) given the raw data on the opponent's strategies,
players select analogy partitions so as to minimize the prediction errors
(either locally or globally). We have highlighted the existence of
environments with self-repelling analogy partitions in which some mixing
over analogy partitions must arise as well as environments with
self-attractive analogy partitions for which multiple analogy partitions can
arise in equilibrium. In the case of environments with self-repelling
analogy partitions, an implication of our analysis is that faced with the
same objective datasets and the same objective constraints (as measured by
the number of classes), players must be processing information in a
heterogeneous way in equilibrium. Our derivation of this insight follows
from the strategic nature of the interaction, and it should be contrasted
with other possible motives of heterogeneity, for example, based on the
complexity of processing rich datasets.\footnote{%
Such forms of heterogeneity are implicitly suggested in Aragones et al
(2005) (when they highlight that finding regularities in complex datasets is
NP-hard) or Sims (2003) \ (who develops a rational inattention perspective
to model agents who would be exposed to complex environments).} The case of
self-attractive analogy partitions suggests a novel source of multiplicity,
not discussed in the previous literature, and it may shed light on why, in
such environments, different societies may settle down on different
behaviors and different models of expectation formation. Our applications
reveal among other things that environments with strategic complements have
self-attractive analogy partitions and environments with strategic
substitutes have self-repelling analogy partitions.

This study can be viewed as a first step toward a more complete
understanding of the structure and impact of categorizations on expectation
formation in strategic interactions. Among the many possible future research
avenues, one could consider the effect of allowing the clustering to be
based not only on the opponent's distribution of action but also on other
characteristics of the interaction such as the own payoff structure. One
could also consider the effect of allowing the pool of subjects assigned to
the role of a given player to be heterogeneous in the number of categories
that can be considered.

\section*{Appendix}

\textbf{Proof of Proposition 1}

The column player partitions games finely. Let row player's analogy
partition be $An_{1}=\{\{x_{1},x_{2}\},\{x_{3}\}\}$ with $x_{1}<x_{2}$, and
let $\beta _{L}$ denote the probability attached to $L$ in the expectation $%
\beta _{1}(x_{1},x_{2})$.

In any ABEE, the Nash equilibrium strategies are played in $x_{3}$, thus the
column player plays $L$ with probability $\frac{1}{2+x_{3}}$ in $x_{3}$ to
make the row player indifferent. The row player must be mixing in game $%
x_{1} $ too: if $\sigma _{1}(x_{1})=D$, then $\sigma _{2}(x_{1})=R$ which
implies $\beta _{L}\geq \frac{1}{2}$, while the best response of the row
player in $x_{1}$, given $\beta _{L}$, would be $U$; If $\sigma
_{1}(x_{1})=U $ is a best response for the row player, it has to be that $%
\beta _{L}\geq \frac{1}{2+x_{1}}$ which is greater than $\frac{1}{2+x_{2}}$
and so the row player would play $U$ in $x_{2}$ as well. This in turn would
imply that column player would play $R$ in both games, which would violate
the consistency requirement on $\beta _{L}$.

Therefore, any ABEE requires $\beta _{L}=\frac{1}{2+x_{1}}$. Given $\beta
_{L}$, row plays $\sigma _{1}(x_{2})=U$ leading column to play $\sigma
_{2}(x_{2})=R$. By consistency, the column player must be playing $L$ with
probability $\frac{2}{2+x_{1}}$ in game $x_{1}$.

Note that the strategy profile $\sigma _{2}$ just obtained is such that $%
An_{1}$ is not locally clustered. If $x_{1}=b$, the probability of playing $%
L $ in $x_{1}$ is greater than $\frac{1}{2}$, while $\beta _{L}=\frac{1}{2+b}%
<\frac{1}{2+a}<\frac{1}{2}$, where $\frac{1}{2+a}$ is the probability
assigned to $L$ by row's expectations in $x_{3}=a$: game $x_{1}$ should be
reassigned to $x_{3}$. Whenever $x_{1}=a$, regardless of $x_{2}$ being $b$
or $c$, $x_{2}$ should be reassigned to $x_{3}$ because the probability of $%
L $ being played in $x_{2}$ is zero, which is closer to $\frac{1}{2+x_{3}}$
than to $\beta _{L}=\frac{1}{2+a}$. \textbf{Q.E.D}

\medskip \noindent\textbf{Proof of Theorem 1.}

\noindent Compared to classic existence results in game theory, the main
novelty is to show that the global clustering correspondence has properties
that allow to apply Kakutani fixed point theorem to a grand mapping $%
M:\Sigma\times \Lambda \rightrightarrows \Sigma\times \Lambda $, which is a
composition of the following functions and correspondences. Given $%
(\sigma,\lambda)$, we aggregate the strategies following (2) obtaining $\Bar{%
\sigma}$ and we define $\beta$ to be consistent with $\Bar{\sigma}$ for all
analogy partitions. We denote this function $AG$. We perform global
clustering ($GC$) on $(\bar{\sigma},\beta)$ and obtain all $\lambda^{\prime
} $ with support only over partitions that are globally clustered with
respect to $\bar{\sigma}$. Also, starting from $(\bar{\sigma},\beta)$, the
Best Response correspondence ($BR$) yields the optimal strategies for each
analogy partition. Then, we obtain the following composition:

\vspace{-4mm} 
\begin{equation*}
(\sigma,\lambda ) \mapsto _{AG} (\bar{\sigma},\beta )\mapsto _{BR\times GC}
(\sigma^{\prime},\lambda ^{\prime })
\end{equation*}%
here $M(\sigma,\lambda )$ denotes the set of $(\sigma^{\prime
},\lambda^{\prime })$ that can be obtained through this composition. That
is, $M(\sigma,\lambda)=\{(\sigma^{\prime },\lambda^{\prime })\lvert (\bar{%
\sigma},\beta)=AG(\sigma,\lambda),\sigma^{\prime }\in BR(\bar{\sigma}%
,\beta),\lambda^{\prime }\in GC(\bar{\sigma},\beta)\}$

\noindent Note that $AG$ is a continuous functions, while $BR$ and $GC $ are
correspondences. The mapping $BR$ is upper-hemicontinuous (uhc) with
non-empty, convex and compact values by standard arguments. Since, as we
will prove later, $GC$ is also uhc with non-empty, convex and compact value, it follows that:

(i) $M$ is nonempty;

(ii) $M$ is uhc and compact-valued because the correspondence $BR\times GC$
is uhc and compact-valued as product of uhc and compact-valued mappings (see
proposition 11.25 in Border (1985)) and the composition $(BR\times GC \circ
AG) (\sigma,\lambda)$ is uhc as composition of uhc mappings (proposition
11.23 in Border (1985)) and compact-valued since $AG$ is single-valued and $%
BR\times GC$ is compact-valued;

(iii) $M$ is convex-valued since $AG$ is single-valued, while $BR$ and $GC$
are convex-valued. This can be established by noting that $M(\sigma ,\lambda
)=BR(AG(\sigma ,\lambda ))\times GC(AG(\sigma ,\lambda ))$. If $(\sigma
^{1},\lambda ^{1})$ and $(\sigma ^{2},\lambda ^{2})$ are both in $M(\sigma
,\lambda )$, then for all $\gamma \in \lbrack 0,1]$, $\gamma \sigma
^{1}+(1-\gamma )\sigma ^{2}\in BR$ because $BR$ is convex-valued and $\gamma
\lambda ^{1}+(1-\gamma )\lambda ^{2}\in GC$ because $GC$ is convex-valued,
thereby implying that $\gamma (\sigma ^{1},\lambda ^{1})+(1-\gamma )(\sigma
^{2},\lambda ^{2})\in M(\sigma ,\lambda )$. Hence, $M(\sigma ,\lambda )$ is
convex for all $(\sigma ,\lambda )$.

Since $\Sigma\times \Lambda $ is a compact and convex set, properties (i) to
(iii) ensure that $M$ has a fixed point by Kakutani's theorem. A fixed point
is a globally CD-ABEE because $GC\circ AG$ ensure globally clustered
partitions and $BR\circ AG$ ensure D-ABEE strategies.

To conclude the proof we need to show the properties of the $GC$
correspondence. $GC$ maps $\Bar{\Sigma}\times \left( (\Delta A_i)^{K_i|%
\mathcal{K}_i|}\times (\Delta A_j)^{K_j|\mathcal{K}_j|} \right)$ into $%
\Lambda $, where all sets are convex and compact. The image $GC(\bar{\sigma}%
,\beta)$ is given by the following set \vspace{-3mm} 
\begin{equation*}
\lambda^{\prime }\in \{\lambda \in \Lambda |\lambda _{i}(An_{i})>0 \implies
An_{i}\in \arg \min_{An_{i}^{\prime }\in \mathcal{K}_{i}}V(\bar{\sigma}%
_{j},\beta _{i}^{\prime })\}
\end{equation*}

\vspace{-3mm} \noindent where $V_{i}(\bar{\sigma}_{j},\beta _{i}^{\prime
})=\sum_{\alpha _{i}\in An_{i}^{\prime }}p(\alpha _{i})\sum_{\omega \in
\alpha _{i}}p(\omega |\alpha _{i})d\left( \bar{\sigma}_{j}(\omega ),\beta
_{i}(\alpha _{i}|An_{i}^{\prime })\right) $.

The latter set can be decomposed in $G_{i}(\bar{\sigma}_{j})$ and $G_{j}(%
\bar{\sigma}_{i})$. Note that $G_{i}$ is nonempty because $\mathcal{K}_{i}$
is finite. Also, $G_{i}$ is a simplex hence it is convex and compact. Thus, $%
GC$ is nonempty, convex and compact valued. We check that $GC(\bar{\sigma}%
,\beta )$ is uhc by showing that it has a closed graph.

We first establish the continuity of $V_{i}$ by verifying that $d$ is a
continuous function. When $d$ is the squared Euclidean distance, $d$ is
clearly continuous in $\bar{\sigma}$ and in $\beta $. Instead, when $d$
represents the KL divergence, it is not generally continuous because
whenever there is $\omega \in \alpha _{i}$ such that $supp[\bar{\sigma}%
_{j}(\omega )]\not\subset supp[\beta _{i}(\alpha _{i}|An_{i})]$, then $d(%
\Bar{\sigma}_{j},\beta _{i})$ goes to infinity. However, the consistency
requirements impose $supp[\bar{\sigma}_{j}(\omega )]\subseteq supp[\beta
_{i}(\alpha _{i}|An_{i})]$. Since global clustering imposes for both players
that $\beta _{i}$ is consistent with $\bar{\sigma}_j$, for all $\omega
,\,\alpha _{i}$ and $An_{i}$, then $d(\Bar{\sigma}_{j},\beta _{i})$ is
finite. Recall that, $d(x,y)=\sum_{a}(x_{a}\ln x_{a}-x_{a}\ln y_{a})$. Since 
$x_{a},y_{a}\in \lbrack 0,1]$ and $x_{a}>0$ implies $y_{a}>0$, under the
convention that $0\ln 0=0$, $d$ is continuous when it represents the KL
divergence. Hence, $V_{i}$ is continuous, if $\beta _{i}$ is consistent with 
$\Bar{\sigma}_{j}$ for both players.

We can now proceed to establish that $GC$ is uhc. Assume by contradiction
that $GC$ does not have a closed graph. That is, $\bar{\sigma}%
_{j}^{n}\rightarrow \bar{\sigma}_{j}$, $\lambda _{i}^{n}\rightarrow \lambda
_{i}$ and $\lambda _{i}^{n}\in G_{i}(\bar{\sigma}_{j}^{n})$, but $\lambda
_{i}\notin G_{i}(\bar{\sigma}_{j})$. Note that $\lambda _{i}\notin G_{i}(%
\bar{\sigma}_{j})$ implies that there is some $\tilde{An}_{i}\in
supp[\lambda _{i}]$ and $\varepsilon ^{\prime },\varepsilon >0$ such that 
\vspace{-1mm} 
\begin{equation*}
+\infty >V_{i}(\bar{\sigma}_{j},\tilde{\beta}_{i})\geq V_{i}(\bar{\sigma}%
_{j},\beta _{i})+\varepsilon +\varepsilon ^{\prime }\text{,}
\end{equation*}%
where $\tilde{\beta}_{i}$ is consistent with $\bar{\sigma}_{j}$ according to 
$\tilde{An}_{i}$. Also, let $\tilde{\beta}_{i}^{n}$ be consistent with $\bar{%
\sigma}_{j}^{n}$, according to $\tilde{An}_{i}$. We want to show that for
some $n$, $\lambda _{i}^{n}(\tilde{An}_{i})>0$ and $V_{i}(\bar{\sigma}%
_{j}^{n},\tilde{\beta}_{i}^{n})>V_{i}(\bar{\sigma}_{j}^{n},\beta _{i}^{n})$.
For $\lambda _{i}^{n}\rightarrow \lambda _{i}$ and $\lambda _{i}(\tilde{An}%
_{i})>0$, for any $n$ large enough, $\lambda _{i}^{n}(\tilde{An}_{i})>0$. As 
$\bar{\sigma}_{j}^{n}\rightarrow \bar{\sigma}_{j}$, for $n$ large enough, by
continuity of $V_{i}$, $V_{i}(\bar{\sigma}_{j}^{n},\beta _{i}^{n})$ is in a
neighborhood of $V_{i}(\bar{\sigma}_{j},\beta _{i})$ so we can write: $V_{i}(%
\bar{\sigma}_{j},\beta _{i})>V_{i}(\bar{\sigma}_{j}^{n},\beta
_{i}^{n})-\varepsilon $. Then, $V_{i}(\bar{\sigma}_{j},\tilde{\beta}%
_{i})\geq V_{i}(\bar{\sigma}_{j},\beta _{i})+\varepsilon +\varepsilon
^{\prime }>V_{i}(\bar{\sigma}_{j}^{n},\beta _{i}^{n})+\varepsilon ^{\prime }$%
. Similarly, $\bar{\sigma}_{j}^{n}\rightarrow \bar{\sigma}_{j}$ implies
that, for any $n$ large enough, $V_{i}(\bar{\sigma}_{j}^{n},\tilde{\beta}%
_{i}^{n})>V_{i}(\bar{\sigma}_{j},\tilde{\beta}_{i})-\varepsilon ^{\prime }$.
Thus,

\vspace{-5mm} 
\begin{equation*}
V_{i}(\bar{\sigma}_{j}^{n},\tilde{\beta}_{i}^{n})>V_{i}(\bar{\sigma}_{j},%
\tilde{\beta}_{i})-\varepsilon ^{\prime }\geq V_{i}(\bar{\sigma}_{j},\beta
_{i})+\varepsilon >V_{i}(\bar{\sigma}_{j}^{n},\beta _{i}^{n})
\end{equation*}

We get $V_{i}(\bar{\sigma}_{j}^{n},\tilde{\beta}_{i}^{n})>V_{i}(\bar{\sigma}%
_{j}^{n},\beta _{i}^{n})$ and $\lambda _{i}^{n}(\tilde{An}_{i})>0$, which
contradicts $\lambda _{i}^{n}\in G_{i}(\bar{\sigma}_{j}^{n})$. It follows
that $GC$ is uhc. \textbf{Q.E.D.}

\medskip\noindent\textbf{Proof of Propositions 8}

Consider the increasing sequence $\{\mu_k\}_{k=0}^K$ with $\mu_0=0$, $\mu_k=\frac{E[\mu|(\mu_{k-1},\mu_k]]+E[\mu|(\mu_{k},%
\mu_{k+1}]]}{2}$, and $\mu_K=1 $.\footnote{%
We prove the existence of such a sequence in Lemma \ref{equid} in the Online
Appendix B.} We check that the sequence we propose satisfies the conditions
for local clustering.

That is, $An$ is locally clustered if, for all $k=1,\dots,K-1$, (i) $%
\left(\beta(\alpha_k)-a(\mu_k|\alpha_k) \right)^2 \leq \left(
\beta(\alpha_{k+1})-a(\mu_k|\alpha_k) \right)^2$, and (ii) $%
\left(\beta(\alpha_{k+1})-a(\mu_k|\alpha_{k+1}) \right)^2 \leq \left(
\beta(\alpha_{k})-a(\mu_k|\alpha_{k+1}) \right)^2$.

It is readily verified that condition (i) reduces to $E[\mu|(\mu_{k},%
\mu_{k+1}]]\geq\frac{2\mu_k-E[\mu|(\mu_{k-1},\mu_{k}]]}{1+2C(\mu_k-E[\mu|(%
\mu_{k-1},\mu_{k}]])}$. By substituting $\mu_k$ in the inequality with $%
\frac{E[\mu|(\mu_{k-1},\mu_{k}]]+E[\mu|(\mu_{k},\mu_{k+1}]]}{2}$ we obtain:

\vspace{-4mm} 
\begin{equation*}
E[\mu|(\mu_{k},\mu_{k+1}]]>\frac{E[\mu|(\mu_{k},\mu_{k+1}]]}{%
1+2CE[\mu|(\mu_{k},\mu_{k+1}]]}
\end{equation*}

\vspace{-1mm} \noindent which is true because the denominator is greater
than 1.

Condition (ii) is equivalent to requiring that, if $E[\mu |(\mu _{k-1},\mu
_{k}]]<\frac{1}{2C}$, the following inequality holds: $E[\mu |(\mu _{k},\mu
_{k+1}]]\leq \mu _{k}+\frac{\mu _{k}-E[\mu |(\mu _{k-1},\mu _{k}]]}{%
1-2CE[\mu |(\mu _{k-1},\mu _{k}]]}$. Recalling that $\mu _{k}=\frac{E[\mu
|(\mu _{k-1},\mu _{k}]]+E[\mu |(\mu _{k},\mu _{k+1}]]}{2}$, we obtain

\vspace{-4mm} 
\begin{equation*}
\frac{E[\mu |(\mu _{k},\mu _{k+1}]]-E[\mu |(\mu _{k-1},\mu _{k}]]}{2}< \frac{%
\frac{E[\mu |(\mu _{k},\mu _{k+1}]]-E[\mu |(\mu _{k-1},\mu _{k}]]}{2}}{%
1-2CE[\mu |(\mu _{k-1},\mu _{k}]]}
\end{equation*}

\vspace{-1mm} \noindent which holds because $0<1-2CE[\mu |(\mu _{k-1},\mu
_{k}]]<1$ when $E[\mu |(\mu _{k-1},\mu _{k}]]<\frac{1}{2C}$.

Since both conditions hold strictly, sequences in the neighborhood of $\{\mu
_{k}\}_{k=0}^{K}$ would also satisfy the conditions for local clustering. 
\textbf{Q.E.D.}

\newpage\setcounter{page}{1}

\section*{For Online Publication}

\subsection*{Online Appendix A: Learning model 1 (Raw data and Machine
Learning)}

In this part of the appendix we formalize the first model of learning
environment introduced in section \ref{learning}.

\subparagraph{Learning dynamics.}

There is a continuum of mass 1 of subjects assigned to the role of player $%
i=1,2$.

We introduce two perturbations that are used to deal with possible
indifferences. First, when playing a game, we assume the payoffs are
slightly perturbed (see Fudenberg and Kreps (1993) or Esponda and Pouzo
(2016)). Specifically, let $\widetilde{\rho }_{i}$ be a random variable with
a continuous density $g_{i}$ on $\left[ 0,1\right] $ and $\varepsilon >0$ a
number that should be thought of as small ($\varepsilon $ measures the
degree of perturbation)and we assume that the distribution of realizations
in the population matches the densities and probabilities induced by $p$ and 
$\widetilde{\rho }_{i}$. Player $i$ in game $\omega $ (with draws $\rho
_{i}(a_{i},\omega ),\rho _{i}(a_{i}^{\prime },\omega )$) picks action $a_{i}$
whenever for all $a_{i}^{\prime }\neq a_{i}$,\footnote{%
Cases of indifference are insignificant whenever $\widetilde{\rho }$ is
distributed in the continuum as assumed here.}%
\begin{equation*}
u_{i}(a_{i},\beta _{i},\omega )+\varepsilon \rho _{i}(a_{i},\omega
)>u_{i}(a_{i}^{\prime },\beta _{i},\omega )+\varepsilon \rho
_{i}(a_{i}^{\prime },\omega )
\end{equation*}%
where $\beta _{i}$ refers to player $i$'s expectation about player $j$'s
behavior in $\omega $.

The second perturbation concerns how clustering is implemented. Suppose in
the previous period $\overline{\sigma }_{j}(\omega )$ represents the
aggregate play of population $j$ when playing game $\omega $. Let $%
\widetilde{\eta }_{i}$ be a random vector with continuous density $h_{i}$
over the interior of $\Delta A_{j}$. We assume that a given player of
population $i$ implements a (global) clustering into $K_{i}$ classes of $%
\left( \overline{s}_{j}(\omega )\equiv \frac{\overline{\sigma }_{j}(\omega
)+\varepsilon \eta _{i}(\omega )}{1+\varepsilon }\right) _{\omega \in \Omega
}$ where $\eta _{i}(\omega )$ is a realization drawn from $\widetilde{\eta }%
_{i}$ and the draws are assumed to be independent across games $\omega $. As
before, we assume that the distributions of realizations in the population
match the density $\widetilde{\eta }_{i}$. To be more specific, player $i$
(with draws $\eta _{i}(\omega )$) picks the partitioning into $K_{i}$
classes so as to solve\footnote{%
For generic $\eta _{i}(\omega )$, there is a unique solution, thus the
handling of indifferences is inconsequential when $\widetilde{\eta }_{i}$
has a density with no atom, as assumed here.}

\begin{equation*}
\arg \min_{P_{i}\in \mathcal{K}_{i}}{\displaystyle\sum_{c_{i}\in P_{i}}}%
p(c_{i}){\displaystyle\sum_{\omega \in c_{i}}}p(\omega \mid c_{i})d(%
\overline{s}_{j}(\omega ),\beta _{i}(c_{i})),
\end{equation*}%
note that in cluster $c_{i}$, the belief $\beta _{i}$ is identified with the
mean of $\overline{s}_{j}(\omega )$ conditional on $\omega \in c_{i}$.

The learning dynamics is as described in the main text and it is fully
pinned down by the initial values of $\overline{\sigma }^{0}(\omega )$ used
in period 1 (as well as $\varepsilon $, $g_{i}$, $h_{i}$).

\subparagraph{Steady state.}

In this part, we show that for a fixed $\varepsilon $, there always exists a
steady state of the learning dynamics just described. We then show that the
limits of such steady states as $\varepsilon $ converges to $0$ correspond
to the globally clustered distributional ABEE.

\begin{proposition}
For a fixed $\varepsilon $, there always exists a steady state of the
learning dynamics.
\end{proposition}

\textbf{Proof.} The proof shares similarities with the purification
techniques introduced by Harsanyi (1973). We consider the same grand mapping 
$M$ that we introduced in the proof of Theorem 1, but now in the perturbed
environment. The perturbations of the payoffs make best-responses
single-valued as commonly observed in the previous learning literature. The
main novelty here is that the perturbations on the strategies at the
clustering stage make the clustering mapping single-valued too. The argument
to show this is a bit more involved than for the payoff perturbation part
because the perturbations at the clustering stage do not allow for additive
separability.

More precisely, consider the compound mapping $(\Bar{\sigma},\lambda
)\mapsto _{C}(\beta ,\lambda )\mapsto _{BR}(\sigma ^{\prime },\lambda
)\mapsto _{AG}(\Bar{\sigma}^{\prime },\lambda )$ where $\Bar{\sigma}^{\prime
}$ is the profile of aggregate best-responses, given $\lambda $.

Fix the probability distributions over analogy partitions $\lambda $. From
the profile of aggregate strategies $\Bar{\sigma}=(\Bar{\sigma}_{1},\Bar{%
\sigma}_{2})$ we can compute the corresponding analogy-based expectations $%
(\beta _{i}(\cdot |An_{i}))_{An_{i}\in supp\lambda _{i}}$ that are
consistent with $\Bar{\sigma}$. The mapping $(\Bar{\sigma},\lambda )\mapsto
(\beta ,\lambda )$ is continuous, single-valued and defined over convex and
compact sets.

We consider best-responses in the perturbed environment. Let us order the
actions in $A_{i}$, so that $a_{i}^{z}$ is the $z$-th element in $A_{i}$. We
denote by $a_{i}^{\ast}(\omega |An_{i})(\cdot )$ the function that maps each
realization of the profile of random variables $\Tilde{\rho_{i}}$ for each
action $a_{i}\in A_{i}$ in game $\omega$ into a best-response, and we write $%
a_{i}=a_{i}^{\ast }(\omega |An_{i})(\rho _{i})$ to indicate that $a_{i}$ is
played when the profile of realizations is $\rho _{i}(\omega )=(\rho
_{i}(a_{i}^{\prime },\omega))_{a_{i}^{\prime }}$ where $\rho _{i}=(\rho
_{i}(\omega ))_{\omega }$. We denote by $X_{i}^{z}(a_{i}^{\ast
}(\omega|An_{i}))$ the set of perturbations under which the action $%
a_{i}^{z} $ is chosen according to $a_{i}^{\ast }(\omega |An_{i})$. That is:

\vspace{-5mm} 
\begin{equation}
X_{i}^{z}(a_{i}^{\ast }(\omega |An_{i}))=\{\rho _{i}|a_{i}^{z}=a_{i}^{\ast
}(\omega |An_{i})(\rho _{i})\}.  \label{DEFX}
\end{equation}
The mixed strategy played by player $i$, under the analogy partition $An_{i}$
in game $\omega $ is induced by $a_{i}^{\ast }(\omega |An_{i})$ if and only
if $\sigma _{i}(\omega|An_{i})$ assigns probability $p_{i}(a_{i}^{z};\omega
,An_{i})$ to $a_{i}^{z}$ where

\vspace{-4mm} 
\begin{equation}
p_{i}(a_{i}^{z};\omega ,An_{i})=\int \dots \int_{\rho _{i}\in
X_{i}^{z}(a_{i}^{\ast }(\omega |An_{i}))}d\rho _{i}(a_{i}^{1},\omega )\dots
d\rho _{i}(a_{i}^{|A_{i}|},\omega )g_{i}(\rho _{i}(a_{i}^{1},\omega ))\dots
g_{i}(\rho _{i}(a_{i}^{|A_{i}|},\omega ))  \label{PROBA}
\end{equation}%
and $g_{i}(\rho _{i})$ is the continuously differentiable pdf of $\Tilde{%
\rho _{i}}$.

Consider first the mapping $BR:(\beta ,\lambda )\mapsto (\sigma ^{\prime
},\lambda )$, where $\sigma ^{\prime }$ is a profile of mixed strategies
that is a best-response to $\beta $. Let $a_{i}^{\ast }(\omega |An_{i})$
prescribe actions that are best responses to $\beta _{i}(\cdot |An_{i})$,
given the perturbed payoffs. Then $BR$ is single-valued (because the set of
realizations of the perturbations under which there are indifferences has
measure zero), and it is readily verified that $BR$ is a continuous function
over convex and compact sets.

Consider the $AG$ function, $(\sigma ^{\prime },\lambda )\mapsto (\Bar{\sigma%
}^{\prime },\beta^{\prime })$ which aggregates the strategies over games and
computes consistent expectations. $AG$ is single-valued and continuous.
Thus, the compound mapping $(\Bar{\sigma},\lambda )\mapsto _{C}(\beta
,\lambda )\mapsto _{BR}(\sigma ^{\prime },\lambda )\mapsto _{AG}(\Bar{\sigma}%
^{\prime },\beta^{\prime })$ is also continuous and single valued over
convex and compact sets.

For the clustering part, the argument is somewhat similar to the best
response part. Consider the clustering mapping $GC:(\Bar{\sigma}^{\prime
},\beta ^{\prime })\mapsto (\Bar{\sigma}^{\prime },\lambda ^{\prime })$,
where $\lambda ^{\prime }=(\lambda _{i}^{\prime },\lambda _{j}^{\prime })$
is such that $\lambda _{i}^{\prime }$ solves the global clustering problem
for player $i$.

Consider the perturbed strategies $\Bar{s}$, where $\Bar{s}_{j}(\omega )=%
\frac{\Bar{\sigma}_{j}(\omega )+\varepsilon \eta _{i}(\omega )}{%
1+\varepsilon }$ and impose that $\beta _{i}$ is consistent with $\Bar{s}%
_{j} $. As established in Theorem 1, for each $\omega \in \alpha $, the
function $d(\Bar{s}_{j}(\omega),\beta _{i}(\alpha |An_{i}))$ is continuous
in $\Bar{s}_{j}$.

We define $An_{i}^{\ast }(\eta _{i})$ as the function mapping the
realization of the perturbation $\Tilde{\eta}_{i}$ to an analogy partition $%
An_{i}$ that solves the clustering problem. As before, we denote by $%
X_{i}^{k}(An_{i}^{\ast })=\{\eta _{i}|An_{i}^{k}=An_{i}^{\ast }(\eta _{i})\}$
the set of realizations such that the $k$-th analogy partition is prescribed
by $An_{i}^{\ast }$.

The mixture of analogy partitions $\lambda _{i}$ is induced by $An_{i}^{\ast
}(\cdot )$ iff $\lambda _{i}$ assigns probability $q(An_{i}^{k})$ to $%
An_{i}^{k}$ where $q(An_{i}^{k})=\int \dots \int_{\eta _{i}\in
X_{i}^{k}(An_{i}^{\ast })}d\eta _{i}(\omega _{1})\dots d\eta _{i}(\omega
_{N})h_{i}(\eta _{i}(\omega _{1}))....h_{i}(\eta _{i}(\omega _{N}))$, where $%
h_{i}$ is the continuously differentiable pdf of $\eta _{i}$. We show now
that the clustering mapping is single-valued. To establish this, we rely on
results from chapter 2 in Milnor (1965). More precisely, we show that if $%
An_{i}$ and $An_{i}^{\prime }$ yield the same $V$ value (the criterion used
for the clustering problem), then the set of realizations of $\Tilde{\eta
_{i}}$ that allow this has measure zero. Given, $\Bar{\sigma}_{j}$ we can
define the function $h(\eta _{i})=V_{i}(\Bar{s}_{j},\beta _{i}(\cdot
|An_{i}))-V_{i}(\Bar{s}_{j},\beta _{i}(\cdot |An_{i}^{\prime }))$, which is
a mapping $h:U\rightarrow R$, where $\eta _{i}\in U$.\footnote{$V_{i}$ is
defined as in the proof of Theorem 1.} The function $h$ is smooth (all
partial derivatives exist and are continuous). Since $\eta _{i}(a_{i},\omega
)>0$, for all $\omega $ and all $a_{i}$, then $U$ is an open set. As $h(\eta
_{i})=0$ is a regular value,\footnote{%
To show this, we note that the first derivatives of $h(\cdot )$ wrt to $\eta
_{i}(a_{i},\omega )$ are linearly independent as one varies $a_{i}$ and $%
\omega $.} then the set $\{\hat{\eta}_{i}|h(\hat{\eta}_{i})=0\}$ is a smooth
manifold of dimension $dim(U)-1=(|A_{i}|-1)\cdot |\Omega |-1$, which has
measure zero in $U$. Then, the argument for $C$ being single valued and
continuous are the same as those used for $BR$.

Thus, the compound mapping

\vspace{-4mm} 
\begin{equation*}
M:(\Bar{\sigma},\lambda )\mapsto_C (\beta ,\lambda )\mapsto_{BR} (\sigma
^{\prime },\lambda )\mapsto_{AG} (\Bar{\sigma}^{\prime },\beta^{\prime
})\mapsto_{GC} (\Bar{\sigma}^{\prime },\lambda ^{\prime })
\end{equation*}

is single-valued and continuous, and it maps $\bar{\Sigma}\times \Lambda $
into $\bar{\Sigma}\times \Lambda $, which are convex and compact sets. By
Brouwer's fixed point theorem, this mapping has a fixed point.

It is then readily verified that the fixed point $(\sigma ,\lambda )$ is a
steady state of the learning dynamics. \textbf{Q.E.D.}

\begin{proposition}
Consider a sequence of steady states $(\sigma ^{(\varepsilon )},\lambda
^{(\varepsilon )})$ of the learning dynamics induced by $\varepsilon $ where 
$\sigma ^{(\varepsilon )}$ denotes the ex ante strategy (prior to the
realizations of the perturbations $\rho $) and $\lambda ^{(\varepsilon )}$
denotes the distribution of the profile of analogy partitions.\footnote{$%
\sigma _{i}^{(\varepsilon )}$ is a strategy of player $i$ that depends on
the game $\omega $ and the analogy partition $An_{i}$ of player $i$ as in
the general construction above.} Consider an accumulation point $(\sigma
,\lambda )$ of $(\sigma ^{(\varepsilon )},\lambda ^{(\varepsilon )})$ as $%
\varepsilon $ tends to $0$. $(\sigma ,\lambda )$ is a globally clustered
distributional ABEE.
\end{proposition}

\noindent\textbf{Proof.} If $\lim_{\varepsilon\rightarrow
0}(\sigma^{(\varepsilon)},\lambda^{(\varepsilon)})=(\sigma,\lambda)$, then
for $\varepsilon$ small enough $supp[\sigma_i(\omega)]\subseteq
supp[\sigma_i^{(\varepsilon)}(\omega)]$ and $supp[\lambda_i]\subseteq
supp[\lambda_i^{(\varepsilon)}]$, for $i=1,2$ and $\omega\in\Omega$.

Since $(\sigma ^{(\varepsilon )},\lambda ^{(\varepsilon )})$ is a steady
state, any $An_{i}\in supp[\lambda _{i}^{(\varepsilon )}]$ solves the
clustering problem for player $i$ in the perturbed environment. Thus, for $%
\varepsilon =0$, $An_{i}\in supp[\lambda _{i}]$ solves the clustering
problem because $d$ is continuous in $\varepsilon $ (as established in
Theorem 1, imposing consistency on $\beta $ suffices to guarantee continuity
in the case of KL divergence). The same argument can be made to show that $%
\sigma $ is a best-response to $\lambda $. Thus, $(\sigma ,\lambda )$ is a
steady state of the learning dynamics when $\varepsilon =0$.

It follows that $(\sigma ,\lambda )$ is a globally clustered ABEE because
the requirements for the equilibrium and the steady states coincide when $%
\varepsilon =0$ and the independence of the random draws ensures that $%
\sigma \in \Sigma _{1}\times \Sigma _{2}$ and $\lambda \in \Delta\mathcal{K}
_{1}\times \Delta\mathcal{K} _{2}$. \textbf{Q.E.D.}


\subsection*{Online Appendix B}

\textbf{Globally CD-ABEE in Example 1}

We also provide a description of the globally clustered distributional ABEE.

Let $An_x=\{\{x\},\{x^{\prime },x^{\prime \prime }\}\}$ denote an analogy
partition of player 1 and $\lambda_x=Pr(An_x)$. Let $p_x$ denote the
probability that $L$ is played in game $x$ by player 2.

Let $p_{x}\leq p_{x^{\prime }}\leq p_{x^{\prime \prime }}$. For the purposes
of global clustering it must be the case that $p_{x^{\prime }}=\frac{%
p_{x}+p_{x^{\prime \prime }}}{2}$, and also $\lambda _{x^{\prime }}=0$
unless $p_{x}=p_{x^{\prime }}=p_{x^{\prime \prime }}$, but it is readily
verified that there is no ABEE such that $p_{x}=p_{x^{\prime }}=p_{x^{\prime
\prime }}$.

Then, in order to ensure global clustering, player 2's strategies must be
such that $p_{x}<p_{x^{\prime }}=\frac{p_{x}+p_{x^{\prime \prime }}}{2}%
<p_{x^{\prime \prime }}$. It is easily verified that there is no
distributional ABEE where player 2 is playing pure strategies in game $x$ or 
$x^{\prime \prime }$. Thus, player 2 is mixing in both games and, clearly,
also in $x^{\prime }$. For player 2's indifference, the aggregate strategy
of player 1 $\bar{\sigma}_{1}(x)$ must be to play $U$ with probability $%
\frac{1}{2}$ in all games $x$. In order to sustain such aggregate strategies
for player 1, whenever $x<x^{\prime }<x^{\prime \prime }$ player 1 must be
indifferent in game $x^{\prime \prime }$ when using $An_{x}$ and in game $x$
when using $An_{x^{\prime \prime }}$.

To illustrate this, take $x<x^{\prime }<x^{\prime \prime }$. Player 1
expects $L$ to be played with probability $\frac{p_{x}+3p_{x^{\prime \prime
}}}{4}$ in games $\{x^{\prime },x^{\prime \prime }\}$ when using $An_{x}$,
and with probability $\frac{3p_{x}+p_{x^{\prime \prime }}}{4}$ in games $%
\{x,x^{\prime }\}$ when using $An_{x^{\prime \prime }}$. By setting the
former probability equal to $\frac{1}{2+x}$ and the latter equal to $\frac{1%
}{2+x^{\prime \prime }}$, we can always sustain $\bar{\sigma}_{1}(x)$ and $%
\bar{\sigma}_{1}(x^{\prime \prime })$ where $U$ is played with probability $%
\frac{1}{2}$. Also, since $\frac{p_{x}+3p_{x^{\prime \prime }}}{4}<\frac{1}{%
2+x^{\prime }}<\frac{3p_{x}+p_{x^{\prime \prime }}}{4}$, player 1 plays $U$
in $x^{\prime }$ when using $An_{x}$ and $D$ when using $An_{x^{\prime
\prime }}$. Therefore, $U$ is played in aggregate in $x^{\prime }$ with
probability $\frac{1}{2}$ if and only if $\lambda _{x}=\lambda _{x^{\prime
\prime }}=\frac{1}{2}$.\bigskip

\vspace{\baselineskip} \textbf{ABEE in Beauty-Contest game, section 3.1}

We provide here the details needed to compute the ABEE in the framework
introduced in section 3.1, in the context of the Beauty-Contest game
illustration. We assume players use the same analogy partition $%
(\Theta_k)_{k=1}^K$. For $\theta\in\Theta_k$, player $i$'s best response is: 
$a_i(\theta)=(1-r)\theta+r E[a_j(\theta)|\Theta_k]$.

Then, the average best response of $i$ in $\theta$ is: 
\begin{equation*}
E[a_i(\theta)|\Theta_k]=\int_{\Theta_k}a_i(\theta)f(\theta|\Theta_k)d(%
\theta)=(1-r)E[\theta|\Theta_k]+r E[a_j(\theta)|\Theta_k]\iff
\end{equation*}
\begin{equation*}
E[a_i(\theta)|\Theta_k]=(1-r)E[\theta|\Theta_k]+r \left(
(1-r)E[\theta|\Theta_k]+r E[a_i(\theta)|\Theta_k] \right)
\end{equation*}
and so both players expects their opponent to play the average $\theta$ in
the analogy class: $E[a_1(\theta)|\Theta_k]=E[a_2(\theta)|\Theta_k]=E[%
\theta|\Theta_k]$ and they play: 
\begin{equation*}
a_1^{ABEE}(\theta)=a_2^{ABEE}(\theta)=(1-r)\theta+rE[\theta|\Theta_k]
\end{equation*}

\bigskip \textbf{Proof of Proposition 7.}

\noindent By definition, $\sigma_j(\mu)\in BR_j(\mu,\beta_i(\alpha_k))$
implies that the following equations must hold in equilibrium:

\vspace{-10mm} 
\begin{equation*}
\sigma_j(\mu)=A+\mu B+\mu C\int_{\mu_{k-1}}^{\mu_k} \frac{f(\mu)}{%
F(\mu_k)-F(\mu_{k-1})}\sigma_i(\mu)d\mu \hspace{45mm}
\end{equation*}

\vspace{-4mm} 
\begin{equation*}
=A+\mu\left(B+AC+BC\,E[\mu|\alpha_k]+
C^2E[\mu|\alpha_k]\int_{\mu_{k-1}}^{\mu_k}\frac{f(\nu)}{F(\mu_k)-F(\mu_{k-1})%
}\sigma_j(\nu)d\nu\right)
\end{equation*}

Taking the weighted average of $\sigma _{j}(\mu )$ over the interval $[\mu
_{k-1},\mu _{k}]$, according to the distribution of $\mu $, yields the
following equation:

\vspace{-4mm} {\footnotesize 
\begin{equation*}
\int_{\mu_{k-1}}^{\mu_k} \frac{f(\mu)}{F(\mu_k)-F(\mu_{k-1})}%
\sigma_j(\mu)d\mu=
\end{equation*}
\begin{equation*}
=A+E[\mu|\alpha_k]\left(B+AC+BC\,E[\mu|\alpha_k]+C^2E[\mu|\alpha_k]\int_{%
\mu_{k-1}}^{\mu_k} \frac{f(\nu)}{F(\mu_k)-F(\mu_{k-1})}\sigma_j(\nu)d\nu%
\right)
\end{equation*}
}

\vspace{-3mm} \noindent By consistency of $\beta _{i}(\alpha _{k})$, the
equation above simplifies into $\beta _{i}(\alpha _{k})=\frac{A+B\,E [\mu
|\alpha _{k}]}{1-C\,E[\mu |\alpha _{k}]}$.

In equilibrium, both players have the same expectations $\beta _{1}(\alpha
_{k})=\beta _{2}(\alpha _{k})$. Substituting the expression of $\beta
_{i}(\alpha _{k})$ into the best-responses yields the following equilibrium
(pure) strategies: for all $\alpha _{k}\in An_{1}$ (and $An_{2}$) and for
all $\mu \in \alpha _{k}$,

\vspace{-4mm} 
\begin{equation*}
\sigma_1(\mu)=\sigma_2(\mu)=A+\mu \frac{B+AC}{1-C\,E[\mu|\alpha_k]}
\end{equation*}

This is the unique ABEE and it is symmetric. \textbf{Q.E.D.}

\bigskip

\begin{lemma}
\label{Loc_implies_glob} Let $\sigma $ be some strategy profile and $d$ be
either the squared Euclidean distance or the KL divergence. If $A n_{i} \in 
\mathcal{K}_{i}$ is a globally clustered analogy partition for player $i$
with respect to $\sigma $, then $A n_{i}$ is a locally clustered analogy
partition for player $i$ with respect to $\sigma $.
\end{lemma}

\noindent\textbf{Proof.} Let $A n_{i}$ be a globally clustered analogy
partition with respect to $\sigma $. Let $\beta _{i}$ be consistent with $%
\sigma $. Assume by contradiction that $A n_{i}$ is not locally clustered.
\noindent Then, $\exists \alpha _{i} ,\alpha \prime _{i} \in A n_{i}\,
\wedge \,\hat{\omega } \in \alpha _{i}$ s.t. $d (\sigma _{j} (\hat{\omega })
,\beta _{i} (\alpha _{i})) >d (\sigma _{j} (\hat{\omega }) ,\beta _{i}
(\alpha \prime _{i}))$. \newline
\noindent Let $\widehat{\alpha }_{i} =\alpha _{i} \setminus \{\hat{\omega }%
\} $ and $\widehat{\alpha \prime }_{i} =\alpha \prime _{i} \cup \{\hat{%
\omega }\}$. Then, {\small \vspace{-1mm}%
\begin{equation*}
\sum _{\omega \in \alpha _{i}}p (\omega ) d (\sigma _{j} (\omega ) ,\beta
_{i} (\alpha _{i})) +\sum _{\omega \in \alpha \prime _{i}}p (\omega ) d
(\sigma _{j} (\omega ) ,\beta _{i} (\alpha \prime _{i}))
\end{equation*}%
} {\small \vspace{-1mm}%
\begin{equation*}
>\sum _{\omega \in \widehat{\alpha _{i}}}p (\omega ) d (\sigma _{j} (\omega
) ,\beta _{i} (\alpha _{i})) +\sum _{\omega \in \widehat{\alpha \prime _{i}}%
}p (\omega ) d (\sigma _{j} (\omega ) ,\beta _{i} (\alpha \prime _{i}))
\end{equation*}%
}{\small 
\begin{equation*}
=p (\widehat{\alpha _{i}}) \sum _{\omega \in \widehat{\alpha _{i}}}p (\omega
\vert \widehat{\alpha _{i}})d (\sigma _{j} (\omega ) ,\beta _{i} (\alpha
_{i})) +p (\widehat{\alpha \prime _{i}}) \sum _{\omega \in \widehat{\alpha
\prime _{i}}}p (\omega \vert \widehat{\alpha \prime _{i}})d (\sigma _{j}
(\omega ) ,\beta _{i} (\alpha \prime _{i}))
\end{equation*}%
}

{\small 
\begin{equation*}
>p (\widehat{\alpha _{i}}) \sum _{\omega \in \widehat{\alpha _{i}}}p (\omega
\vert \widehat{\alpha _{i}})d (\sigma _{j} (\omega ) ,\beta _{i} (\widehat{%
\alpha _{i}})) +p (\widehat{\alpha \prime _{i}}) \sum _{\omega \in \widehat{%
\alpha \prime _{i}}}p (\omega \vert \widehat{\alpha \prime _{i}})d (\sigma
_{j} (\omega ) ,\beta _{i} (\widehat{\alpha \prime _{i}}))
\end{equation*}%
} \noindent where the second inequality holds because, when $\beta_i$ is
consistent with the strategies played in the games in a given analogy class,
this is the representative object that minimizes the sum of prediction
errors in that analogy class among all possible representative objects in $%
\Delta A_j$.

\noindent Let $\widehat{A n}_{i} =\widehat{\alpha }_{i} \cup \widehat{\alpha
\prime }_{i} \cup \{A n_{i} \setminus \{\alpha _{i} ,\alpha \prime _{i}\}\}$%
, then: 
\begin{equation*}
\sum _{\alpha _{i} \in A n_{i}}p (\alpha _{i}) \sum _{\omega \in \alpha
_{i}}p (\omega \vert \alpha )d (\sigma _{j} (\omega ) ,\beta _{i} (\alpha
_{i})) >\sum _{\alpha _{i} \in \widehat{A n}_{i}}p (\alpha _{i}) \sum
_{\omega \in \alpha _{i}}p (\omega \vert \alpha _{i})d (\sigma _{j} (\omega
) ,\beta _{i} (\alpha _{i}))
\end{equation*}

\noindent which contradicts $A n_{i}$ being a globally clustered analogy
partition. \textbf{Q.E.D.}

\vspace{\baselineskip }

\begin{lemma}
\label{truncation} (Reverse Truncation) Let $X$ be a RV on $[0,1]$ with
continuous pdf $f_{X}$ and cdf $F_{X}$. There exists a continuous pdf $g$
over $[0,+\infty )$, such that: 
\begin{equation*}
g_{X|[0,1]}(x)=f_{X}(x),\,\text{where}\,\,\,g_{X|[0,1]}(x)=\frac{g_X(x)}{%
Pr_g[0\leq X\leq 1]}
\end{equation*}
\end{lemma}

\noindent\textbf{Proof.} We want to find a continuous function $g_X$ such
that: 
\begin{equation*}
g_X(x)= 
\begin{cases}
Pr_g[0\leq X\leq 1]f_X(x) & 0\leq x\leq 1 \\ 
v(x) & 1 < x < \infty%
\end{cases}%
\end{equation*}
Note that $g_X$ is continuous if $v(\cdot)$ is continuous and $v(1)=
Pr_g[0\leq X\leq 1]f_X(1)$. Also, $g_X$ must be a pdf, so it must be the
case that $\int_0^{+\infty}g_X(x)dx=1$. That is, $\int_0^1Pr_g[0\leq X\leq
1]f_X(x)dx+\int_1^{+\infty}v(x)dx=Pr_g[0\leq X\leq 1]+(1-Pr_g[0\leq X\leq
1])=1$.

\noindent Let us pick the right function $v(.)$. This function must be
continuous and satisfy two conditions: (i) $\int_1^{+\infty}v(x)dx=1-T$, and
(ii) $v(1)= Tf_X(1)$, where $T\equiv Pr_g[0\leq X\leq 1]$. Let $v()$ be
defined as $v(x)=Tf(1)e^{\frac{Tf(1)}{1-T}(1-x)}$ which is continuous since,
for $a,b\in\mathbb{R}$, the function $a e^{bx}$ is continuous in $x$. Also, $%
v(1)=Tf(1)e^{\frac{Tf(1)}{1-T}(0)}=Tf(1)$.

And finally: $\int_1^{+\infty}v(x)dx= Tf(1)e^{\frac{Tf(1)}{1-T}}
\int_1^{+\infty}e^{-\frac{Tf(1)}{1-T}x} dx = 1-T$. \textbf{Q.E.D.}

\vspace{\baselineskip}

\begin{lemma}
\label{equid} There always exists an equidistant-expectations partition.
\end{lemma}

\noindent\textbf{Proof.} We show that there always exists a sequence $\{\mu
_{k}\}_{k=0}^{k}$ with $\mu _{0}=0 $, some $\mu _{1}\in \lbrack 0,1]$ and,
for $k=2,\dots ,K$, $\mu _{k}$ defined so that $\mathbb{E}[\mu|(\mu
_{k-1},\mu _{k}]]=2\mu _{k-1}-\mathbb{E}[\mu |(\mu _{k-2},\mu _{k-1}]]$. We
note that if $\mu _{1}$ is too large, $\mu _{K}$ might be above 1. Lemma \ref%
{truncation} allows us to consider $\mu $ to be a random variable from $%
[0,+\infty )$ distributed according to a continuous strictly positive pdf $g$%
, and cdf $G$, with $g(\mu )=f(\mu )G(1)$, for $0\leq \mu \leq 1$.

Let $\mu _{k}\geq \mu _{k-1}\geq 0$. Since $g(\mu )$ is strictly positive
and continuous in $\mu $, then $\mathbb{E}[\mu |(\mu _{k-1},\mu _{k}]]=\frac{%
1}{G(\mu _{k})-G(\mu _{k-1})}\int_{\mu _{k-1}}^{\mu _{k}}\mu g(\mu )d\mu $
is continuous and strictly decreasing in $\mu _{k-1}$, and it is continuous
and strictly increasing in $\mu _{k}$. Moreover, if $\mu _{k}\leq 1$, we
have:

\vspace{-4mm} 
\begin{equation*}
\frac{\int_{\mu_{k-1}}^{\mu_{k}}\mu g(\mu)d\mu}{G(\mu_{k})-G(\mu_{k-1})} = 
\frac{\int_{\mu_{k-1}}^{\mu_{k}}\mu G(1)f(\mu)d\mu}{G(1)(F(\mu_{k})-F(%
\mu_{k-1}))}
\end{equation*}
so the term $G(1)$ cancels out and we are back to the original distribution $%
F$.

Fix $\mu_{k-1}$. The function $m(\mu_{k})\equiv\mathbb{E}[\mu|[\mu_{k-1},%
\mu_{k}]]$ is continuous and strictly increasing over $(\mu_{k-1},+\infty)$,
with image $(\mu_{k-1},+\infty)$. Then the inverse function $m^{-1}$ exists
over $(\mu_{k-1},+\infty)$ and it is continuous and strictly increasing over 
$(\mu_{k-1},+\infty)$.

Given $\mu _{k-2},\mu _{k-1}$, we use the inverse function to retrieve $\mu
_{k}$ from the equation $\mathbb{E}[\mu |(\mu _{k-1},\mu _{k}]]=2\mu _{k-1}-%
\mathbb{E}[\mu |(\mu _{k-2},\mu _{k-1}]]$. Let $h(\mu _{k-2},\mu
_{k-1})\equiv 2\mu _{k-1}-\mathbb{E}[\mu |(\mu _{k-2},\mu _{k-1}]]$. Note
that $h(\cdot )$ is a continuous function and $h(\mu _{k-2},\mu _{k-1})\geq
\mu _{k-1}$.

Starting from $\mu _{0}=0$ and some $\mu _{1}(\mu _{1})\equiv \mu _{1}$, we
recursively define $\mu _{k}$ as a function of $\mu_{1}$ as follows: $\mu
_{2}(\mu _{1})=m^{-1}(h(\mu _{0},\mu _{1}))$ and \vspace{-3mm} 
\begin{equation*}
\mu _{k}(\mu _{1})=m^{-1}(h(\mu _{k-2}(\mu _{1}),\mu _{k-1}(\mu _{1})))
\end{equation*}

\vspace{-3mm} \noindent for $k=3,\dots ,K$. Note that, for each $k$, the
function $\mu _{k}(\mu _{1}) $ is well defined. Since $h(\mu _{k-2},\mu
_{k-1})\geq \mu _{k-1}$, then the inverse function exists at the point $%
h(\mu _{k-2},\mu _{k-1})$.

Since $m^{-1}:(\mu _{k-1},+\infty )\rightarrow (\mu _{k-1},+\infty )$ is
also strictly increasing and continuous, then $\mu _{k}(\mu _{1})$ is
continuous, being a composition of continuous functions, and $\mu _{k}(\mu
_{1})\geq \mu _{k-1}$, with equality if and only if $h(\mu _{k-2},\mu
_{k-1})=\mu _{k-1}\iff \mu _{k-1}=\mathbb{E}[\mu |(\mu _{k-2},\mu
_{k-1}]]\iff \mu _{k-1}=\mu _{k-2}$. So, either we get the sequence with 0
everywhere, or a strictly increasing sequence.

Let $\mu _{1}=0$, then $\mu _{K}(0)=0$. Let $\mu _{1}=1$, then $\mu
_{K}(1)>1 $. Then, by the intermediate value theorem, there must exist $%
0<\mu _{1}^{\ast }<1$ such that $\mu _{K}=1$. \textbf{Q.E.D.}

\end{document}